\documentclass[groupedaddress,reprint,amsmath,amssymb,aps,prb,floatfix]{revtex4-1}

\usepackage{graphicx}
\usepackage{dcolumn}
\usepackage{bm}
\usepackage{physics}
\usepackage[colorlinks,
    linkcolor=blue,
    anchorcolor=red,
    citecolor=green
]{hyperref}

\usepackage{simpler-wick}
\usepackage{array}
\usepackage{cancel}
\usepackage{times}
\usepackage[caption = false]{subfig}

\newcommand{\PRLSection}[1]{{\textit{\color{blue} #1}.}}

\renewcommand{\v}[1]{\mathbf{#1}}

\begin{document}

\title{Type-II heavy Fermi liquids and the magnetic memory of 4Hb-TaS$_2$}
\author{Elio J. K\"{o}nig}
\affiliation{Max-Planck-Institut f\"{u}r Festk\"{o}rperforschung, Heisenbergstra{\ss}e 1, 70569 Stuttgart, Germany }

\date{\today}

\begin{abstract}
The interplay of quantum spin liquids with itinerant conduction electrons is of crucial interest for understanding layered structures composed of frustrated magnet and metal monolayers. Using parton-mean-field theory, we here demonstrate that a type-II heavy Fermi liquid, which is characterized by a vortex lattice in the slave boson condensate, can occur in the vicinity of the quantum phase transition separating fractionalized and heavy Fermi liquid phases. The magnetic flux threading each such vortex is about $ v_f/ 137 c$ times smaller than the magnetic flux threading vortices in type-II superconductors, where $v_f$ is the speed of magnetic excitations and $c$ the speed of light. This makes a magnetic observation of this effect challenging. We propose scanning tunneling spectroscopy  instead and investigate its signatures. If a type-II heavy Fermi liquid is cooled into a type-II superconductor, vortices in the slave boson condensate and in the superconducting condensate mutually attract. We argue that the type-II heavy Fermi liquid thereby provides a compelling explanation for the magnetic memory observed recently [Persky \textit{et al.}, Nature \textbf{609}, 692 (2022)] in thermal cycles of 4Hb-TaS$_2$.
\end{abstract}

\maketitle

\PRLSection{Introduction}
Unconventional superconductors emerging in quantum materials with strongly entangled spin degrees of freedom are fascinating phases of matter of prime theoretical and experimental interest. While classic examples are heavy fermion superconductors~\cite{SteglichFranz1979} and cuprates~\cite{BednorzMueller1986}, this interest was recently boosted by the advent of van-der-Waals materials~\cite{GeimGrigorieva2013}. In particular, the past decade witnessed tremendous experimental progress in the ability to synthesize, stack, align, twist, and gate a variety of two-dimensional materials with superconducting, magnetic, semimetallic, and semiconducting properties.

\begin{figure}
\includegraphics[scale=1]{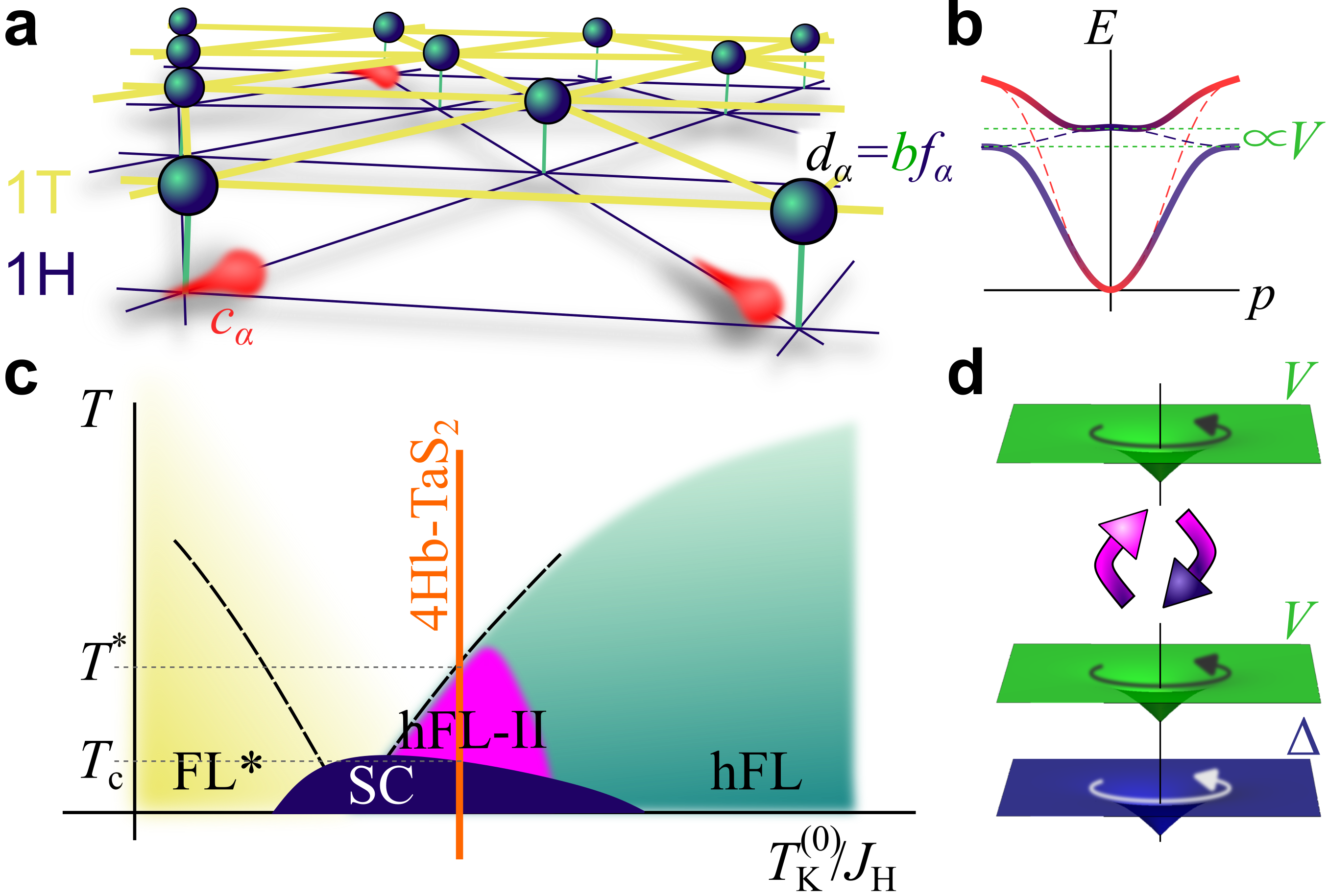}
\caption{a) Triangular superlattice of a 1T-1H bilayer of TaS$_2$, i.e.~the building block of the layered 4Hb bulk. Strong Coulomb repulsion amongst $d$ electrons in the 1T-layer is treated using slave bosons $b$.  b)  Once $\langle b \rangle \propto V$ condenses, heavy $f$ fermions hybridize with light $c$ electrons leading to a heavy Fermi liquid (hFL). c) Schematic phase diagram 
(temperature $T$ versus the ratio of bare Kondo temperature $T_K^{(0)}$ and spin-exchange $J_H$ between adjacent 1T-sites). The quantum critical point between U(1) fractionalized Fermi liquid (FL$^*$, $b$ uncondensed) and heavy Fermi liquid (hFL, $b$ condensed) is buried under a superconducting (SC) dome close to a type-II hFL phase (hFL-II) displaying vortices
in $V$ upon application of external magnetic fields. d) Mechanism for magnetic memory: Below $T_c$, vortices in $V$ and in the SC order parameter $\Delta$ attract. Upon heating to temperatures $T \in [T_c, T_*]$, SC order disappears. Part of the vortices in $V$ remains pinned even when the external magnetic field is turned off and 
nucleates SC vortices in subsequent cooling cycles.}
\label{fig:Summary}
\end{figure}

This letter is motivated by experiments on allotropes of the  transition metal dichalcogenide TaS$_2$, which are all layered and exfoliable. 
2H-TaS$_2$ is a metal turning into an Ising superconductor below $T_c = 0.7 K$~\cite{NagataTsutsumi1992,DeLaBarreraHunt2018}. In contrast, 1T-TaS$_2$ forms a $\sqrt{13}\times \sqrt{13}$ charge density wave at about 200K with a typical ``star-of-David'' deformation of the Ta atoms. Below this temperature, the electronic degrees of freedom are commonly described by a single-orbital Hubbard model on the emergent triangular superlattice, where the distance between the centers of the unit cell ensures relatively flat bands in comparison with the interaction scale. This leads to Mott localization~\cite{FazekasTosatti1979,RibakKanigel2017,LutsykKlusek2018,MuruyamaMatsuda2020} (cf.~\cite{VanoLiljeroth2021,AyanideParga2022,QiaoLiu2017,ChenCrommie2020,RuanCrommie2021,WanUgeda2022,ChenCrommie2022,LiuShih2021} for related materials). 
Fascinatingly, 1T-TaS$_2$ seems to avoid magnetic order down to lowest accessible temperatures~\cite{KratochvilovaPark2017,RibakKanigel2017,Klanjvsek2017} and given thermodynamic evidence for low-lying (fermionic) excitations~\cite{RibakKanigel2017,MuruyamaMatsuda2020} it was proposed~\cite{LawLee2017,HeLee2018} that 1T-TaS$_2$ realizes a (gapless) quantum spin liquid (QSL)~\cite{SavaryBalents2016,ZhouNg2017}. Finally, 4Hb-TaS$_2$, composed of alternating 1T-TaS$_2$ and 1H-TaS$_2$ layers, is an unconventional superconductor with enhanced $T_c$ = 2.7K (as compared to 2H-TaS$_2$)~\cite{RibakKanigel2020,NayakBeidenkopf2021,SilberDagan2022,AlmoalemKanigel2022} in which the $\mu$SR signal has a drastic change at $T_c$, consistent with time-reversal symmetry breaking (TRSB) concomitant with the onset of superconductivity. 

\begin{table*}
    \centering
    \begin{tabular}{l|c|c|c|c}
        theory & TRSB above $T_c$ & carrier of magnetic memory & topological order & generalizable to mixed-valence\\
        \hline \hline
        chiral spin liquid~\cite{PerskyKalisky2022} & yes & chiral order parameter $\chi_\triangle$ & yes: U(1)  \cite{footnoteCSL} & no \\
        \hline
        spinon-vortex interconversion~\cite{Lin2022} & yes & flux in $\v a$, trapped spinons
 & no & yes\\ \hline
        vison-vortex interconversion~\cite{Chen2023} &  no & vison & yes: $\mathbb Z_2$ & no\\
        \hline
        type-II heavy Fermi liquid [\textit{this work}]& no & vortex in hybridization $V$ & no &yes
    \end{tabular}
    \caption{Comparison of theories of the magnetic memory in 4Hb-TaS$_2$, as outlined in the main text. }
    \label{tab:Comparison}
\end{table*}

A major puzzle relates to the magnetic memory effect~\cite{PerskyKalisky2022} of 4Hb-TaS$_2$. In this experiment, first, the sample is field-cooled below $T_c$, then the resulting
vortex configuration, which is rather irregular, is mapped out using the scanning SQUID technique. Subsequently, the sample is heated to a temperature $T_{\rm cycle} > T_c$, before being zero-field cooled (ZFC) below $T_c$, again. The magnetic memory consists in the fact that a fraction of vortices (up to 12 \%) survives the cycle, as long as $T_{\rm cycle} < T_* \simeq 3.6$K. At the same time, the measured magnetic field in the normal state is virtually absent, which is consistent with previously mentioned $\mu$SR data~\cite{RibakKanigel2020}. Thermodynamic evidence for spontaneous TRSB above $T_c$ is not reported in the literature. Therefore, a vestigial~\cite{FernandesSchmalian2019} TRSB normal phase~\cite{BojesenSudbo2013,BojesenSudbo2014}, similar to the one recently reported~\cite{GrinenkoBabaev2021} in strongly hole-doped Ba$_{1-x}$K$_x$Fe$_2$As$_2$, is likely not responsible for the magnetic memory. An alternative scenario based on point-like impurities alone (e.g. residual local magnetic moments) seems to be implausible, as point-defects are energetically unable to nucleate entire vortex lines threading the 3D bulk crystals.

\PRLSection{Theories of 4Hb-TaS$_2$} It is a topic of present debate whether 4Hb-TaS$_2$ should be viewed as a Kondo lattice system. While Kondo physics was clearly demonstrated in tunneling spectroscopy of 1T-TaS$_2$ monolayers on 1H-TaS$_2$~\cite{VanoLiljeroth2021}, first principle theoretical calculations~\cite{CrippaValenti2023} highlight the importance of charge transfer from 1T to 1H layers, as spectroscopically observed in 1T monolayers on 4Hb~\cite{WenYan2021}. This leads to a ``mixed-valence picture'' at intermediate charge transfer.  
While ARPES data suggests that the dispersion of the individual constituent layers of 4Hb-TaS$_2$ is retained~\cite{RibakKanigel2020}, very recent STM experiments unveil both charge transfer and Kondo physics in distinct spatial areas on the surface of 4Hb-TaS$_2$~\cite{NayakBeidenkopf2023}.

We briefly review existing theories of the magnetic memory effect, see Tab.~\ref{tab:Comparison} for comparison.
Beena Kalisky and co-authors of the experimental discovery paper~\cite{PerskyKalisky2022}, suggest that the magnetic memory originates from a chiral spin liquid (CSL), which is a gapped QSL with spontaneous TRSB with a non-zero pseudo-scalar order parameter $\chi_\triangle = \vec S_1 \cdot (\vec S_2 \times \vec S_3)$ of spins $\vec S_{1,2,3}$ on the corners of the triangles of the superlattice. This scenario is at odds with the experimental report of TRSB at low $T \leq T_c$, only~\cite{RibakKanigel2020}. To reconcile the contradiction it was argued that the measurable magnetization generated by the CSL is small. 

The details of vortex nucleation in the thermal cycling process within an extended CSL scenario were elaborated subsequently by Shi-Zeng Lin~\cite{Lin2022}, 
who proposes the condensation of the Kondo hybridization $V$, in addition to a non-zero $\chi_\triangle$. At this point, the internal gauge theory is Higgsed, i.e. confining, and thus topological order is lost~\cite{footnoteTopOrder}. Assuming that $V$ is homogeneous, flux-lines in the physical U(1) field $\v A$ imply flux-lines in the emergent U(1) field $\v a$~\cite{ColemanSchofield2005,ColemanBook}, which can trap spinons above $T_c$. These spinons subsequently nucleate Abrikosov vortices upon successive cooling below $T_c$. 

Gang Chen~\cite{Chen2023} proposes an alternative to the CSL scenario, i.e. that the magnetic memory originates from a vison-vortex interconversion due to a postulated underlying $\mathbb Z_2$ topological order of local moments in the normal state.
Expanding on earlier ideas by Senthil and Fisher~\cite{SenthilFisher2001}, vortices in the physical superconducting order parameter $\Delta$ bind fluxes in the emergent $\mathbb Z_2$ gauge field (``visons''). When heated above $T_c$, the emergent fluxes remain and nucleate spontaneous formation of vortices in $\Delta$ in the next cooling cycle. This scenario does not assume TRSB above $T_c$, yet the sign of the vortices reappearing during ZFC cycles is only weakly imposed by fields in the environment.

\PRLSection{Type-II heavy Fermi liquids} Here we consider an alternative scenario for the magnetic memory effect, based on a phase we call ``type-II heavy Fermi liquids (hFLs)''. 
In full analogy to type-II superconductors, it relies on vortices in the complex mean-field order parameter $V$ of hFLs, Fig.~\ref{fig:Summary} d), which denotes the Kondo hybridization in Kondo lattice systems or the condensate wave function of slave-bosons in theories of mixed-valence systems~\cite{ColemanBook}, respectively. We explicitly demonstrate that below $T_c$ and in a magnetic field, vortices in the superconducting order parameter $\Delta$ and vortices in $V$ mutually attract, thereby stabilizing a vortex lattice of superimposed $\Delta$ and $V$ vortices. Upon heating, $\Delta$ vanishes, yet vortices in $V$ persist as long as the temperature is smaller than the Kondo temperature $T_K$ (identified with $T_*$). Some $V$-vortices may remain pinned once the magnetic field is turned off. Importantly, the physical magnetic field they generate is small and possibly below the threshold of experimental sensitivity. During subsequent cooling cycles and by means of the aforementioned attraction mechanism these vortices in $V$ nucleate vortices in $\Delta$. We first discuss this effect phenomenologically, demonstrate its energetic stability and provide microscopic, justifying calculations.
We then discuss the magnetic and tunneling spectroscopic signatures of this phase.

\PRLSection{Main assumptions of the theory} 
Given that we study magnetic field-induced vortices, the first and main assumption of our theory is the tacit supposition that heavy fermion physics is at play and that orbital field effects dominate over Zeeman effects (``Pauli-limit''), not only for $\Delta$ but also for $V$. The latter assumption is motivated by recent susceptibility measurements~\cite{RibakKanigel2017} uncovering a minute magnetic moment per star of David in 1T-TaS$_2$ and by early theoretical proposals~\cite{FazekasTosatti1979} advocating a negligible $g$ factor in the same material. It is however most strongly motivated by the experimental fact that the fields applied in the magnetic memory experiments~\cite{PerskyKalisky2022} are below 1 mT, while the splitting of the Kondo peak in STM is not observed in 4Hb-TaS$_2$ up to 9T~\cite{NayakBeidenkopf2023} and occurs at about 10T in 1T on 1H bilayers~\cite{VanoLiljeroth2021}. 
The second assumption to stabilize type-II hFLs is the condition of large effective penetration depth $\lambda_K$ as compared to the Kondo coherence length $\xi_K$. This is very much akin to the standard condition in type-II superconductors. 
As we explicitly demonstrate, close to the quantum phase transition (QPT) in Fig.~\ref{fig:Summary} c), this assumption is met by the presence of sufficiently strong ring-exchange terms which occur close to Mott-delocalization. These terms, sometimes argued to stabilize  quantum spin liquid ground states~\cite{CookmeyerMoore2021}, may additionally help to energetically stabilize the vortex lattice in the Kondo hybridization $V$ by translating the applied external magnetic field to a net flux of the emergent gauge field~\cite{Motrunich2006,KoenigKomijani2021}. 

\PRLSection{Overview}
Abrikosov's~\cite{Abrikosov1957} explanation of the Ryabinin-Shubnikov superconducting phase~\cite{RyabininShibnikov1935} with imperfect Mei{\ss}ner effect involves 
vortex lattices in the complex, charge $2e$ order parameter $\Delta \sim \langle c_\uparrow c_\downarrow \rangle$, where $c_\alpha$ is the annihilation operator of a conduction fermion of spin $\alpha$. The condition to stabilize the flux phase in type-II superconductors is a large London penetration depth $\lambda$ as compared to the coherence length $\xi$ (which sets the scale of the vortex core diameter).

Kondo lattice systems are commonly described using a fractionalized representation of spins~\cite{ReadNewns1983,Coleman1983} $\hat{\mathbf S} = f^\dagger_\alpha \boldsymbol{\sigma}_{\alpha \beta} f_\beta$ where $f$ are fermionic spinon operators of emergent gauge charge $e_*$. On the mean-field level, the complex Kondo hybridization $V \sim \sum_\alpha \langle c_\alpha f^\dagger_\alpha \rangle$ [which has charge $(e,-e_*)$] condenses. 
Similarly, in mixed-valence systems a spinless, charge $(e,-e_*)$ slave-boson~\cite{Barnes1976,Coleman1984} $b$ may be invoked to represent the physical fermionic creation operator $d_\alpha = b f_\alpha$ on partially field orbitals with infinite Hubbard repulsion, see Fig.~\ref{fig:Summary} a). Loosely speaking, $b$ annihilates the charge, and the spinon $f_\alpha$ the spin of the local $d_\alpha$ electron. The slave boson condensate $\langle b \rangle \propto V$ is adiabatically connected to the Kondo hybridization $V$ of the local-moment system. The QPT in Fig.~\ref{fig:Summary} separates a fractionalized Fermi liquid~\cite{SenthilVojta2003,SenthilSachdev2004} (FL$^*$, $b$ uncondensed, with unit valence/star of David) from the Higgs phase (condensed $\langle b \rangle \propto V$, hFL)~\cite{ColemanSchofield2005,ColemanBook} with a large Fermi surface. Given the apparent analogy to the superconducting $\Delta$, in this letter we study vortex phases of $V$. As in superconductors, such a phase is stabilized by a large ratio $\kappa_K =\lambda_K/\xi_K$ of effective penetration depth and coherence length.

Real space vortices in the slave boson condensate were first considered in the context of cuprates~\cite{WenZee1989,Sachdev1992,NagaosaLee1992}. 
Topological defects in Kondo lattice systems were also discussed over the years, including vortices bound to visons~\cite{TsvelikColeman2022} in Kondo-Kitaev systems, as well as vortices and skyrmions in more richly structured theories of two-channel Kondo lattices~\cite{WugalterColeman2020,KornjavcaFlint2021,GeKomijani2022}. Most important for this work are the discussion of vortex solutions of the hybridization in Kondo lattice systems \cite{SaremiSenthil2009,SaremiSenthil2011,IimuraHoshino2020,footnoteVortex}, which were applied to antiferromagnetic QPTs. 
Less directly connected to the physics of this letter are vortices of $V$ in momentum space, which occur, e.g., in theories of $\beta$-YbAlB$_4$~\cite{RamiresTsvelik2012}.

\begin{figure}
\includegraphics[scale=1]{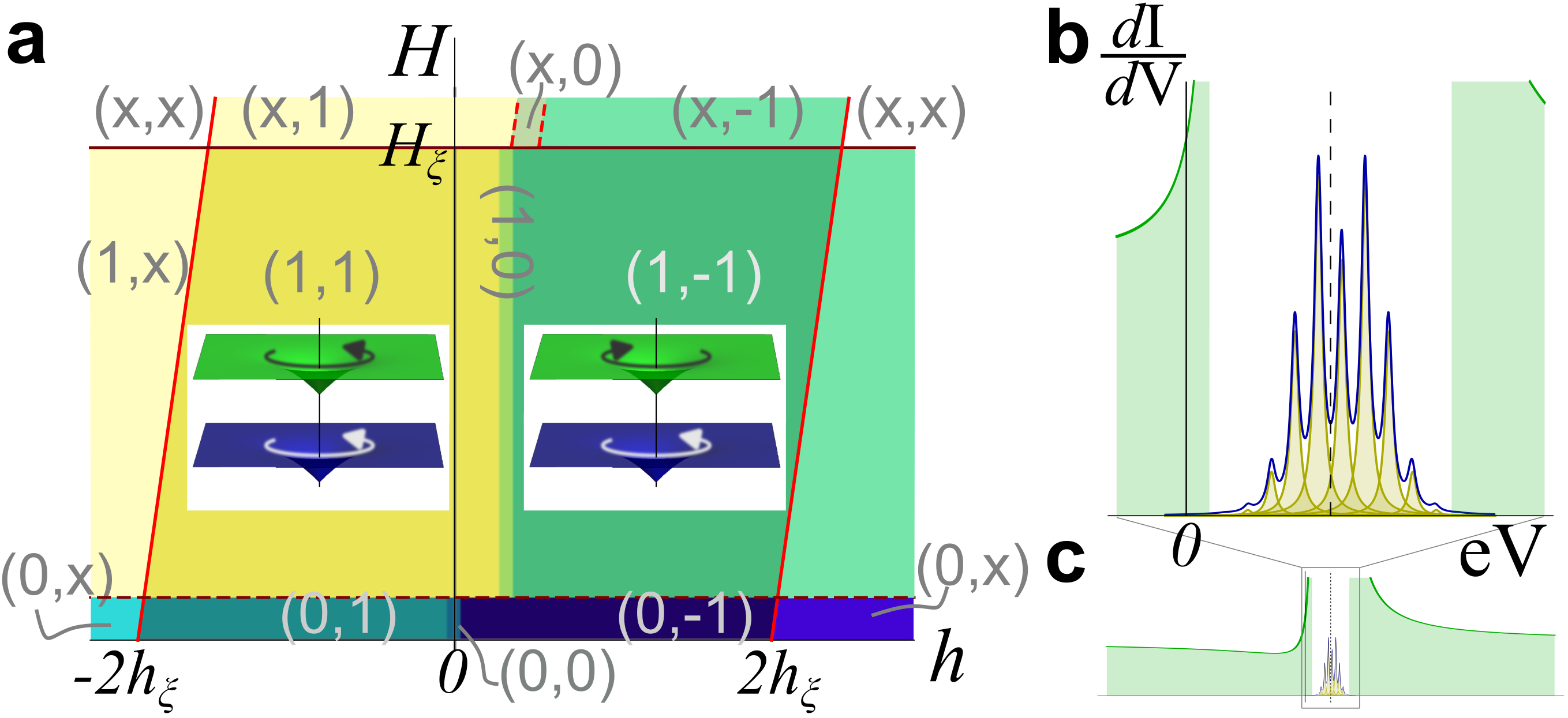}
\caption{a) Phase diagram at fixed $T< T_c$ in the plane spanned by $H$ and $h$. Here, we chose $e^2/e_*^2 = 10^{-4}$, $\xi/ \xi_K = 1.5$, $\kappa = 2 = \kappa_K$. The labels $(\zeta, \zeta_K)$ denote the vorticity if $\zeta, \zeta_K$ is an integer, while the label $x$ denotes the absence of the corresponding order parameter. In most of the parameter space superimposed vortex-vortex/vortex-antivortex are energetically favorable (see illustration in the insets), and these regimes tend to cover all of the parameter space below $H \simeq H_\xi$, $\vert h \vert < 2 h_\xi$  as $\kappa_K, \kappa$ are further increased. b) Tunneling density of states in the center of a Kondo vortex core c) the same at a larger scale.}
\label{fig:MagField}
\end{figure}

\PRLSection{Ginzburg-Landau theory}
The free energy density of the ``order parameters'' and (emergent) gauge fields is $f = f_{\rm SC} + f_{\rm K} + f_{\rm MW} + f_{\rm mw}$, where 
\begin{subequations}
\begin{align}
    f_{\rm SC} & = \alpha \vert \vec \Delta \vert^2 + \beta U(\vec \Delta) + \frac{\vert (- i  \boldsymbol{\nabla} - 2 e \mathbf A) \vec \Delta \vert^2}{2m_{\rm SC}},\label{eq:fSC} \\
    f_{\rm K} & = \alpha_K \vert V \vert^2 + \beta_K \vert V\vert^4 + \frac{\vert (- i  \boldsymbol{\nabla} -  e \mathbf A + e_* \mathbf a) V \vert^2}{2m_{\rm K}}, \label{eq:fK}\\
    f_{\rm MW} & = \frac{\mathbf B^2 -2  \mathbf B\cdot \mathbf H}{8 \pi},\quad f_{\rm mw} = \frac{\mathbf b^2 -2  \mathbf b\cdot \mathbf h}{8 \pi}.
\end{align}
\label{eq:GL}
\end{subequations}
Setting $\hbar = 1$, we define fundamental flux quanta $\Phi_0 = \pi/e$ and $\Phi_K = \pi/e_*$. 
The parameters $\alpha \sim (T-T_c)$ and $\alpha_K \sim (T-T_K$) define the mean-field transitions.
Bold-face vectors live in physical 3D space. The dispersion of both $V$ and $\vec \Delta = (\Delta_{d_{xy}} , \Delta_{d_{x^2 - y^2}})$ may be highly anistropic with predominant 2D kinetics, but for notational simplicity this is disregarded. Note that Eq.~\eqref{eq:GL} is applicable to both Kondo and mixed-valence systems, alike, still we use the subscript $_K$ (for ``Kondo'') throughout. For 4Hb-TaS$_2$, we consider $U(\vec \Delta) = \vert \vec \Delta \vert^4 +  (\vec \Delta^2) (\vec \Delta^2)^* $, which is minimized by $\vec \Delta \propto(1,i)^T$ corresponding to $d + id$ symmetry. It is important to emphasize that the proposed mechanism of magnetic memory is independent of the pairing symmetry. Consistently, in Figs.~\ref{fig:Summary}, \ref{fig:MagField} we omit the vector on $\vec \Delta$, as if s-wave superconductivity was considered.
The electromagnetic terms $f_{\rm MW}$, $f_{\rm mw}$ contain external fields $\v H, \v h$ and are a functional of internal fields $\v B = \boldsymbol{\nabla}\times \mathbf A, \v b = \boldsymbol{\nabla}\times \mathbf a$, where upper (lower) case letters refer to electromagnetic (emergent) gauge fields. 
The parameters of the Ginzburg-Landau functional, Eq.~\eqref{eq:GL}, define coherence length $\xi = 1/\sqrt{2m_{\rm SC} \vert \alpha\vert}, \xi_K = 1/\sqrt{2m_{\rm K} \vert \alpha_K\vert}$ and penetration depth $\lambda = \sqrt{{m_{\rm SC}}/{(2e)^2 \pi \Delta^2}}, \lambda_K = \sqrt{{m_{\rm K}}/{(2e_*)^2 \pi V^2}}$ for both superconducting and heavy fermion ``order parameters'', where mean-field values of $V$, $\Delta$ are implicitly understood. We remark in passing that, for Cooper pair and slave boson condensation alike, true spontaneous symmetry breaking is disallowed by Elitzur's theorem~\cite{Elitzur1975,ColemanBook} and, beyond mean-field, $T_K$ denotes a crossover temperature.


\PRLSection{Microscopics}
While Eq.~\eqref{eq:GL} follows from symmetry arguments alone, important physical properties, including the presence of the type-II hFL, follow from the values of parameters.
Using a mean-field slave boson theory we microscopically derive those values~\cite{SuppMat}.  
We concentrate on the vicinity of the QPT~\cite{SenthilVojta2003,SenthilSachdev2004,Pepin2008,ConsoliVojta2022} in Fig.~\ref{fig:Summary} c) and approach the hFL from the FL$^*$ side, i.e. at temperatures $T\ll t_f$ much below the bandwidth of $f_\alpha$ fermions. To be concrete, we consider a purely 2D motion of an electron-like pocket of $c_\alpha$ electrons which is smaller than the hole-like pocket of $f_\alpha$ fermions (corresponding Fermi momenta $p_c < p_f$), see Fig.~\ref{fig:Summary} b). Both pockets are treated in a parabolic band approximation, and we control the theory in a large-$N$ expansion. On the mean-field level, where $t_f \sim J_H$ is given by the Heisenberg exchange interaction, the emergent gauge coupling $1/e_*^2$ is stabilized by time-reversal-even four-spin interactions (coupling constant $J_4$) while the field $\mathbf h$ is microscopically generated~\cite{Motrunich2006} by the time-reversal-odd three-spin interaction $J_3 \mathbf S_1 \cdot (\mathbf S_2 \times \mathbf S_3)$ around the corners of a triangle of side length $a$. 
The most important result of the microscopic calculations~\cite{SuppMat} is the smallness of $e^2/e_*^2 \sim J_4 a/137 c \ll 1$, with $c$ the speed of light, and the fact that $\kappa_K \sim \sqrt{J_4 t_c}/J_H$, where $t_c$, i.e., the bandwidth of conduction electrons, is large. As $J_4$ approaches $J_H$ from below near Kondo/Mott delocalization, $\kappa_K \gg 1$ and $e^2/e_*^2 \sim v_f/137 c$, where $v_f$ is the speed of $f$-fermions.

\PRLSection{Inhomogeneous field configurations}  In the following, we study the properties of inhomogeneous order parameter solutions and whether they minimize Eq.~\eqref{eq:GL} (see~\cite{SuppMat} for details).
Below the superconducting transition temperature, $T<T_c$ vortices of winding $\zeta$ ($\zeta_K$) may occur in $\Delta$ ($V$) while for $T_c<T$ only vortices in $V$ occur. By means of the minimal coupling, vortices induce a magnetic total flux $\Phi_{\rm tot} $ (total emergent flux $\phi_{\rm tot}$) through the system
\begin{equation}
\left (\begin{array}{c}
        \Phi_{\rm tot} \\
        \phi_{\rm tot}
    \end{array} \right)  = \begin{cases}\left (\begin{array}{c}
        \zeta\Phi_0 \\
        (\zeta - 2\zeta_K) \Phi_K
    \end{array}\right) , & T<T_c,\\
 \frac{2\zeta_K}{1 + (e/e_*)^2} \left ( \begin{array}{c}
           \Phi_0 \frac{e^2}{e_*^2}  \\
          - \Phi_K 
     \end{array}\right), & T_c<T. 
     \end{cases}
\label{eq:Fluxes}
\end{equation}
 For the solution above $T_c$, note that a vortex in $V$ implies a quantized flux of $e \v B - e_* \v b$, while the orthogonal combination $e_* \v B + e \v b$ is blind to the vortex and in the experimentally relevant case $e \ll e_*$, the flux very unevenly distributes between emergent and physical gauge fields. Using microscopic calculations, we find an extremely small magnetic field $B_{\rm max} \lesssim \Phi_0 e^2 \ln(\kappa_K)/(4\pi e_*^2 \lambda_K^2) \sim 10^{-6} (J_H/t_c)$T in the center of $V$-vortices\cite{SuppMat}, where $J_H/t_c \lesssim 10^{-2}$. We remark in this context that the scale at which magnetization is excluded in published scanning SQUID experiments~\cite{PerskyKalisky2022} is of the order of $10^{-7}$ T which is about two orders of magnitude below the measured signal of superconducting vortices.

We next present the energy of a vortex relative to the homogeneous condensate in the limit of large $\kappa_K \gg 1$ and $e_*^2/e^2$ and to leading order in applied fields $H,h$. For $T<T_c$ we further concentrate on the limit of large $\kappa = \lambda/\xi$  and comparable $\lambda, \lambda_K$ (for other limits see Ref.~\cite{SuppMat})

\begin{equation}
\frac{\Delta F}{L} = \begin{cases} \frac{\Phi_K}{4\pi}\left [\zeta_K^2 h_{\xi_K}\frac{\ln(\kappa_K)}{2 \kappa^2_K} -(\zeta -  2\zeta_K)  h \right ] \\
+  \frac{\Phi_0}{4\pi}\left [\zeta^2 H_{\xi} \frac{\ln(\kappa)}{2 \kappa^2} - \zeta H \right ]  , & T< T_c,\\
\frac{\Phi_K}{4\pi}\left [\zeta_K^2 h_{\xi_K}\frac{\ln(\kappa_K)}{2 \kappa^2_K} -  2\zeta_K ( \frac{e}{e_*} H -  h) \right ], &  T_c<T. \end{cases}
\label{eq:FstringMaintext}
\end{equation}
The reference magnetic fields are $H_\xi = {1}/({2 e \xi^2}), h_{\xi_K} = {1}/({2 e_* \xi_K^2})$, they determine $H_{c,2}$ and $h_{c,2}/2$. The $(\zeta,\zeta_K)$ values minimizing Eq.~\eqref{eq:FstringMaintext} are illustrated in Fig.~\ref{fig:MagField} a). Below $T_c$ and at a finite field, non-zero $\zeta = \vert \zeta_K \vert \neq 0$ are favorable, signaling the attraction of distinct vortex types. This concludes the main ingredients for the type-II hFL scenario of magnetic memory.

\PRLSection{Tunneling into type-II hFLs}
As highlighted above, the characteristic inhomogeneous magnetic signal of $V$-vortices in type-II hFLs may be too weak to be experimentally accessible. We therefore consider another probe, namely tunneling spectroscopy/microscopy of hFL vortex cores. 

Qualitatively, the vanishing hybridization function inside the vortex core implies line defects of FL$^*$ character. At the semiclassical level, the tunneling density of states (DOS) inside the vortex core is therefore small as compared to the hFL far away from the vortices (it is given by the bare mass of conduction electrons rather than the heavy mass). Moreover, in the hFL phase, electrons of the tip can tunnel into both $c-$ and $f$-fermions~\cite{MaltsevaColeman2009}. While hFLs display a rich and characteristic tunneling spectrum due to multiple interfering tunneling channels, semiclassically these features disappear in the vortex core. 

Quantum effects in the tunneling spectra of type-II hFL are expected to be strongest at a finite bias voltage corresponding to the energy $E_{\rm gap}$ around which the hybridization gap opens. Here, the DOS vanishes outside the vortex cores (called ``Kondo insulator'' when $E_{\rm gap}$ coincides with the Fermi energy $E_F$), but bound states are expected to be trapped inside the core~\cite{SaremiSenthil2011}. Using the approximation of parabolic bands employed above and generalizing a classical calculation by Caroli \textit{et al.}~\cite{CaroliMatricon1964}, we demonstrate~\cite{SuppMat} the existence of bound states at energies $E_l = l E_{\rm mg}$, where $l \in \mathbb Z + 1/2$. Note that the minigap $E_{\rm mg}  ={ V_{\rm gap}^2} \sqrt{t_c/J_H}/{J_H}$ in terms of the size of the indirect gap $V_{\rm gap} \sim V \sqrt{J_H/t_c}$ contains an additional large ratio of masses $\sqrt{t_c/J_H}$ leading to clearly spaced subgap states. 
We also determine the wave function of these solutions~\cite{SuppMat}. The most important feature is a suppressing factor $\sqrt{J_H/t_c}$ of the $c-$electron component of the wave function as compared to the $f-$ electron component of the wave function. In the tunneling DOS, which is proportional to the wave function squared, this is the quantum manifestation of the small mass associated with the FL$^*$-esque $V = 0$ physics inside the vortex core. 
In Fig.~\ref{fig:MagField}, b), c) we present the tunneling DOS (arbitrary units) in the vortex core superimposed with the signal of scattering states (or, equivalently, tunneling in the absence of a vortex~\cite{MaltsevaColeman2009}). Note the asymmetry of the signal, which is due to unequal roles played by $c$ and $f$ fermions. A sufficiently sharp tip has strong support with the central subgap states, only. Here we use a tip radius $\sim 3a$, $E_{\rm mg}/V_{\rm gap} = 1/10$, and a broadening of peaks $\Gamma = 0.2 E_{\rm mg}$.

Of course, this discussion of tunneling signatures assumes that the picture of the slave boson condensate survives to energies $E_{\rm gap}$ away from the Fermi energy, which is expected to hold if $\vert E_{\rm gap} - E_F \vert \ll T_{\rm K}$. 

\PRLSection{Discussion} 
In summary, we studied the type-II heavy Fermi liquid phase, in which external magnetic fields induce vortex lattices in the slave boson condensate, or equivalently the Kondo hybridization $V \propto \langle b \rangle$. Within mean-field slave-boson calculations this phenomenon is demonstrated to occur in the vicinity of the QPT between fractionalized Fermi liquid and heavy Fermi liquid, provided ring-exchange terms are strong enough. While the magnetic signal of $V$-vortices is strongly suppressed, we propose tunneling microscopy as a means to experimentally access the effect and present the spectrum of subgap states trapped inside the vortex core. We argue that the type-II heavy Fermi liquid provides a compelling scenario for the magnetic memory effect in 4Hb-TaS$_2$, by means of mutual attraction of superconducting vortices and $V$-vortices combined with pinning effects of $V$-vortices.

\begin{acknowledgments}

It is a pleasure to thank H.~Beidenkopf, P.~M.~Bonetti, P.~Coleman, G.~Khaliullin, Y.~Komijani, J.~Motruk, F.~von~Rohr, M.~Scheurer, Q.~Si as well as F.~L\"upke and R.~Mazzilli for useful discussions. This work was partly performed at the Kavli Institute for Theoretical Physics, thereby this research was supported in part by the National Science Foundation under Grant No. NSF PHY-1748958.
\end{acknowledgments}

\bibliography{MagneticMemory}

\begin{thebibliography}{73}%
\makeatletter
\providecommand \@ifxundefined [1]{%
 \@ifx{#1\undefined}
}%
\providecommand \@ifnum [1]{%
 \ifnum #1\expandafter \@firstoftwo
 \else \expandafter \@secondoftwo
 \fi
}%
\providecommand \@ifx [1]{%
 \ifx #1\expandafter \@firstoftwo
 \else \expandafter \@secondoftwo
 \fi
}%
\providecommand \natexlab [1]{#1}%
\providecommand \enquote  [1]{``#1''}%
\providecommand \bibnamefont  [1]{#1}%
\providecommand \bibfnamefont [1]{#1}%
\providecommand \citenamefont [1]{#1}%
\providecommand \href@noop [0]{\@secondoftwo}%
\providecommand \href [0]{\begingroup \@sanitize@url \@href}%
\providecommand \@href[1]{\@@startlink{#1}\@@href}%
\providecommand \@@href[1]{\endgroup#1\@@endlink}%
\providecommand \@sanitize@url [0]{\catcode `\\12\catcode `\$12\catcode
  `\&12\catcode `\#12\catcode `\^12\catcode `\_12\catcode `\%12\relax}%
\providecommand \@@startlink[1]{}%
\providecommand \@@endlink[0]{}%
\providecommand \url  [0]{\begingroup\@sanitize@url \@url }%
\providecommand \@url [1]{\endgroup\@href {#1}{\urlprefix }}%
\providecommand \urlprefix  [0]{URL }%
\providecommand \Eprint [0]{\href }%
\providecommand \doibase [0]{http://dx.doi.org/}%
\providecommand \selectlanguage [0]{\@gobble}%
\providecommand \bibinfo  [0]{\@secondoftwo}%
\providecommand \bibfield  [0]{\@secondoftwo}%
\providecommand \translation [1]{[#1]}%
\providecommand \BibitemOpen [0]{}%
\providecommand \bibitemStop [0]{}%
\providecommand \bibitemNoStop [0]{.\EOS\space}%
\providecommand \EOS [0]{\spacefactor3000\relax}%
\providecommand \BibitemShut  [1]{\csname bibitem#1\endcsname}%
\let\auto@bib@innerbib\@empty
\bibitem [{\citenamefont {Steglich}\ \emph {et~al.}(1979)\citenamefont
  {Steglich}, \citenamefont {Aarts}, \citenamefont {Bredl}, \citenamefont
  {Lieke}, \citenamefont {Meschede}, \citenamefont {Franz},\ and\ \citenamefont
  {Sch\"afer}}]{SteglichFranz1979}%
  \BibitemOpen
  \bibfield  {author} {\bibinfo {author} {\bibfnamefont {F.}~\bibnamefont
  {Steglich}}, \bibinfo {author} {\bibfnamefont {J.}~\bibnamefont {Aarts}},
  \bibinfo {author} {\bibfnamefont {C.~D.}\ \bibnamefont {Bredl}}, \bibinfo
  {author} {\bibfnamefont {W.}~\bibnamefont {Lieke}}, \bibinfo {author}
  {\bibfnamefont {D.}~\bibnamefont {Meschede}}, \bibinfo {author}
  {\bibfnamefont {W.}~\bibnamefont {Franz}}, \ and\ \bibinfo {author}
  {\bibfnamefont {H.}~\bibnamefont {Sch\"afer}},\ }\href {\doibase
  10.1103/PhysRevLett.43.1892} {\bibfield  {journal} {\bibinfo  {journal}
  {Phys. Rev. Lett.}\ }\textbf {\bibinfo {volume} {43}},\ \bibinfo {pages}
  {1892} (\bibinfo {year} {1979})}\BibitemShut {NoStop}%
\bibitem [{\citenamefont {Bednorz}\ and\ \citenamefont
  {M{\"u}ller}(1986)}]{BednorzMueller1986}%
  \BibitemOpen
  \bibfield  {author} {\bibinfo {author} {\bibfnamefont {J.~G.}\ \bibnamefont
  {Bednorz}}\ and\ \bibinfo {author} {\bibfnamefont {K.~A.}\ \bibnamefont
  {M{\"u}ller}},\ }\href@noop {} {\bibfield  {journal} {\bibinfo  {journal}
  {Zeitschrift f{\"u}r Physik B Condensed Matter}\ }\textbf {\bibinfo {volume}
  {64}},\ \bibinfo {pages} {189} (\bibinfo {year} {1986})}\BibitemShut
  {NoStop}%
\bibitem [{\citenamefont {Geim}\ and\ \citenamefont
  {Grigorieva}(2013)}]{GeimGrigorieva2013}%
  \BibitemOpen
  \bibfield  {author} {\bibinfo {author} {\bibfnamefont {A.~K.}\ \bibnamefont
  {Geim}}\ and\ \bibinfo {author} {\bibfnamefont {I.~V.}\ \bibnamefont
  {Grigorieva}},\ }\href {https://www.nature.com/articles/nature12385}
  {\bibfield  {journal} {\bibinfo  {journal} {Nature}\ }\textbf {\bibinfo
  {volume} {499}},\ \bibinfo {pages} {419} (\bibinfo {year}
  {2013})}\BibitemShut {NoStop}%
\bibitem [{\citenamefont {Nagata}\ \emph {et~al.}(1992)\citenamefont {Nagata},
  \citenamefont {Aochi}, \citenamefont {Abe}, \citenamefont {Ebisu},
  \citenamefont {Hagino}, \citenamefont {Seki},\ and\ \citenamefont
  {Tsutsumi}}]{NagataTsutsumi1992}%
  \BibitemOpen
  \bibfield  {author} {\bibinfo {author} {\bibfnamefont {S.}~\bibnamefont
  {Nagata}}, \bibinfo {author} {\bibfnamefont {T.}~\bibnamefont {Aochi}},
  \bibinfo {author} {\bibfnamefont {T.}~\bibnamefont {Abe}}, \bibinfo {author}
  {\bibfnamefont {S.}~\bibnamefont {Ebisu}}, \bibinfo {author} {\bibfnamefont
  {T.}~\bibnamefont {Hagino}}, \bibinfo {author} {\bibfnamefont
  {Y.}~\bibnamefont {Seki}}, \ and\ \bibinfo {author} {\bibfnamefont
  {K.}~\bibnamefont {Tsutsumi}},\ }\href@noop {} {\bibfield  {journal}
  {\bibinfo  {journal} {Journal of Physics and Chemistry of Solids}\ }\textbf
  {\bibinfo {volume} {53}},\ \bibinfo {pages} {1259} (\bibinfo {year}
  {1992})}\BibitemShut {NoStop}%
\bibitem [{\citenamefont {De~la Barrera}\ \emph {et~al.}(2018)\citenamefont
  {De~la Barrera}, \citenamefont {Sinko}, \citenamefont {Gopalan},
  \citenamefont {Sivadas}, \citenamefont {Seyler}, \citenamefont {Watanabe},
  \citenamefont {Taniguchi}, \citenamefont {Tsen}, \citenamefont {Xu},
  \citenamefont {Xiao} \emph {et~al.}}]{DeLaBarreraHunt2018}%
  \BibitemOpen
  \bibfield  {author} {\bibinfo {author} {\bibfnamefont {S.~C.}\ \bibnamefont
  {De~la Barrera}}, \bibinfo {author} {\bibfnamefont {M.~R.}\ \bibnamefont
  {Sinko}}, \bibinfo {author} {\bibfnamefont {D.~P.}\ \bibnamefont {Gopalan}},
  \bibinfo {author} {\bibfnamefont {N.}~\bibnamefont {Sivadas}}, \bibinfo
  {author} {\bibfnamefont {K.~L.}\ \bibnamefont {Seyler}}, \bibinfo {author}
  {\bibfnamefont {K.}~\bibnamefont {Watanabe}}, \bibinfo {author}
  {\bibfnamefont {T.}~\bibnamefont {Taniguchi}}, \bibinfo {author}
  {\bibfnamefont {A.~W.}\ \bibnamefont {Tsen}}, \bibinfo {author}
  {\bibfnamefont {X.}~\bibnamefont {Xu}}, \bibinfo {author} {\bibfnamefont
  {D.}~\bibnamefont {Xiao}},  \emph {et~al.},\ }\href
  {https://www.nature.com/articles/s41467-018-03888-4} {\bibfield  {journal}
  {\bibinfo  {journal} {Nature communications}\ }\textbf {\bibinfo {volume}
  {9}},\ \bibinfo {pages} {1427} (\bibinfo {year} {2018})}\BibitemShut
  {NoStop}%
\bibitem [{\citenamefont {Fazekas}\ and\ \citenamefont
  {Tosatti}(1979)}]{FazekasTosatti1979}%
  \BibitemOpen
  \bibfield  {author} {\bibinfo {author} {\bibfnamefont {P.}~\bibnamefont
  {Fazekas}}\ and\ \bibinfo {author} {\bibfnamefont {E.}~\bibnamefont
  {Tosatti}},\ }\href
  {https://www.tandfonline.com/doi/abs/10.1080/13642817908245359} {\bibfield
  {journal} {\bibinfo  {journal} {Philosophical Magazine B}\ }\textbf {\bibinfo
  {volume} {39}},\ \bibinfo {pages} {229} (\bibinfo {year} {1979})}\BibitemShut
  {NoStop}%
\bibitem [{\citenamefont {Ribak}\ \emph {et~al.}(2017)\citenamefont {Ribak},
  \citenamefont {Silber}, \citenamefont {Baines}, \citenamefont {Chashka},
  \citenamefont {Salman}, \citenamefont {Dagan},\ and\ \citenamefont
  {Kanigel}}]{RibakKanigel2017}%
  \BibitemOpen
  \bibfield  {author} {\bibinfo {author} {\bibfnamefont {A.}~\bibnamefont
  {Ribak}}, \bibinfo {author} {\bibfnamefont {I.}~\bibnamefont {Silber}},
  \bibinfo {author} {\bibfnamefont {C.}~\bibnamefont {Baines}}, \bibinfo
  {author} {\bibfnamefont {K.}~\bibnamefont {Chashka}}, \bibinfo {author}
  {\bibfnamefont {Z.}~\bibnamefont {Salman}}, \bibinfo {author} {\bibfnamefont
  {Y.}~\bibnamefont {Dagan}}, \ and\ \bibinfo {author} {\bibfnamefont
  {A.}~\bibnamefont {Kanigel}},\ }\href {\doibase 10.1103/PhysRevB.96.195131}
  {\bibfield  {journal} {\bibinfo  {journal} {Phys. Rev. B}\ }\textbf {\bibinfo
  {volume} {96}},\ \bibinfo {pages} {195131} (\bibinfo {year}
  {2017})}\BibitemShut {NoStop}%
\bibitem [{\citenamefont {Lutsyk}\ \emph {et~al.}(2018)\citenamefont {Lutsyk},
  \citenamefont {Rogala}, \citenamefont {Dabrowski}, \citenamefont {Krukowski},
  \citenamefont {Kowalczyk}, \citenamefont {Busiakiewicz}, \citenamefont
  {Kowalczyk}, \citenamefont {Lacinska}, \citenamefont {Binder}, \citenamefont
  {Olszowska}, \citenamefont {Kopciuszynski}, \citenamefont {Szalowski},
  \citenamefont {Gmitra}, \citenamefont {Stepniewski}, \citenamefont
  {Jalochowski}, \citenamefont {Kolodziej}, \citenamefont {Wysmolek},\ and\
  \citenamefont {Klusek}}]{LutsykKlusek2018}%
  \BibitemOpen
  \bibfield  {author} {\bibinfo {author} {\bibfnamefont {I.}~\bibnamefont
  {Lutsyk}}, \bibinfo {author} {\bibfnamefont {M.}~\bibnamefont {Rogala}},
  \bibinfo {author} {\bibfnamefont {P.}~\bibnamefont {Dabrowski}}, \bibinfo
  {author} {\bibfnamefont {P.}~\bibnamefont {Krukowski}}, \bibinfo {author}
  {\bibfnamefont {P.~J.}\ \bibnamefont {Kowalczyk}}, \bibinfo {author}
  {\bibfnamefont {A.}~\bibnamefont {Busiakiewicz}}, \bibinfo {author}
  {\bibfnamefont {D.~A.}\ \bibnamefont {Kowalczyk}}, \bibinfo {author}
  {\bibfnamefont {E.}~\bibnamefont {Lacinska}}, \bibinfo {author}
  {\bibfnamefont {J.}~\bibnamefont {Binder}}, \bibinfo {author} {\bibfnamefont
  {N.}~\bibnamefont {Olszowska}}, \bibinfo {author} {\bibfnamefont
  {M.}~\bibnamefont {Kopciuszynski}}, \bibinfo {author} {\bibfnamefont
  {K.}~\bibnamefont {Szalowski}}, \bibinfo {author} {\bibfnamefont
  {M.}~\bibnamefont {Gmitra}}, \bibinfo {author} {\bibfnamefont
  {R.}~\bibnamefont {Stepniewski}}, \bibinfo {author} {\bibfnamefont
  {M.}~\bibnamefont {Jalochowski}}, \bibinfo {author} {\bibfnamefont {J.~J.}\
  \bibnamefont {Kolodziej}}, \bibinfo {author} {\bibfnamefont {A.}~\bibnamefont
  {Wysmolek}}, \ and\ \bibinfo {author} {\bibfnamefont {Z.}~\bibnamefont
  {Klusek}},\ }\href {\doibase 10.1103/PhysRevB.98.195425} {\bibfield
  {journal} {\bibinfo  {journal} {Phys. Rev. B}\ }\textbf {\bibinfo {volume}
  {98}},\ \bibinfo {pages} {195425} (\bibinfo {year} {2018})}\BibitemShut
  {NoStop}%
\bibitem [{\citenamefont {Murayama}\ \emph {et~al.}(2020)\citenamefont
  {Murayama}, \citenamefont {Sato}, \citenamefont {Taniguchi}, \citenamefont
  {Kurihara}, \citenamefont {Xing}, \citenamefont {Huang}, \citenamefont
  {Kasahara}, \citenamefont {Kasahara}, \citenamefont {Kimchi}, \citenamefont
  {Yoshida}, \citenamefont {Iwasa}, \citenamefont {Mizukami}, \citenamefont
  {Shibauchi}, \citenamefont {Konczykowski},\ and\ \citenamefont
  {Matsuda}}]{MuruyamaMatsuda2020}%
  \BibitemOpen
  \bibfield  {author} {\bibinfo {author} {\bibfnamefont {H.}~\bibnamefont
  {Murayama}}, \bibinfo {author} {\bibfnamefont {Y.}~\bibnamefont {Sato}},
  \bibinfo {author} {\bibfnamefont {T.}~\bibnamefont {Taniguchi}}, \bibinfo
  {author} {\bibfnamefont {R.}~\bibnamefont {Kurihara}}, \bibinfo {author}
  {\bibfnamefont {X.~Z.}\ \bibnamefont {Xing}}, \bibinfo {author}
  {\bibfnamefont {W.}~\bibnamefont {Huang}}, \bibinfo {author} {\bibfnamefont
  {S.}~\bibnamefont {Kasahara}}, \bibinfo {author} {\bibfnamefont
  {Y.}~\bibnamefont {Kasahara}}, \bibinfo {author} {\bibfnamefont
  {I.}~\bibnamefont {Kimchi}}, \bibinfo {author} {\bibfnamefont
  {M.}~\bibnamefont {Yoshida}}, \bibinfo {author} {\bibfnamefont
  {Y.}~\bibnamefont {Iwasa}}, \bibinfo {author} {\bibfnamefont
  {Y.}~\bibnamefont {Mizukami}}, \bibinfo {author} {\bibfnamefont
  {T.}~\bibnamefont {Shibauchi}}, \bibinfo {author} {\bibfnamefont
  {M.}~\bibnamefont {Konczykowski}}, \ and\ \bibinfo {author} {\bibfnamefont
  {Y.}~\bibnamefont {Matsuda}},\ }\href {\doibase
  10.1103/PhysRevResearch.2.013099} {\bibfield  {journal} {\bibinfo  {journal}
  {Phys. Rev. Research}\ }\textbf {\bibinfo {volume} {2}},\ \bibinfo {pages}
  {013099} (\bibinfo {year} {2020})}\BibitemShut {NoStop}%
\bibitem [{\citenamefont {Vano}\ \emph {et~al.}(2021)\citenamefont {Vano},
  \citenamefont {Amini}, \citenamefont {Ganguli}, \citenamefont {Chen},
  \citenamefont {Lado}, \citenamefont {Kezilebieke},\ and\ \citenamefont
  {Liljeroth}}]{VanoLiljeroth2021}%
  \BibitemOpen
  \bibfield  {author} {\bibinfo {author} {\bibfnamefont {V.}~\bibnamefont
  {Vano}}, \bibinfo {author} {\bibfnamefont {M.}~\bibnamefont {Amini}},
  \bibinfo {author} {\bibfnamefont {S.~C.}\ \bibnamefont {Ganguli}}, \bibinfo
  {author} {\bibfnamefont {G.}~\bibnamefont {Chen}}, \bibinfo {author}
  {\bibfnamefont {J.~L.}\ \bibnamefont {Lado}}, \bibinfo {author}
  {\bibfnamefont {S.}~\bibnamefont {Kezilebieke}}, \ and\ \bibinfo {author}
  {\bibfnamefont {P.}~\bibnamefont {Liljeroth}},\ }\href
  {https://www.nature.com/articles/s41586-021-04021-0} {\bibfield  {journal}
  {\bibinfo  {journal} {Nature}\ }\textbf {\bibinfo {volume} {599}},\ \bibinfo
  {pages} {582} (\bibinfo {year} {2021})}\BibitemShut {NoStop}%
\bibitem [{\citenamefont {Ayani}\ \emph {et~al.}(2022)\citenamefont {Ayani},
  \citenamefont {Pisarra}, \citenamefont {Ibarburu}, \citenamefont {Garnica},
  \citenamefont {Miranda}, \citenamefont {Calleja}, \citenamefont
  {Mart{\'\i}n},\ and\ \citenamefont {de~Parga}}]{AyanideParga2022}%
  \BibitemOpen
  \bibfield  {author} {\bibinfo {author} {\bibfnamefont {C.~G.}\ \bibnamefont
  {Ayani}}, \bibinfo {author} {\bibfnamefont {M.}~\bibnamefont {Pisarra}},
  \bibinfo {author} {\bibfnamefont {I.~M.}\ \bibnamefont {Ibarburu}}, \bibinfo
  {author} {\bibfnamefont {M.}~\bibnamefont {Garnica}}, \bibinfo {author}
  {\bibfnamefont {R.}~\bibnamefont {Miranda}}, \bibinfo {author} {\bibfnamefont
  {F.}~\bibnamefont {Calleja}}, \bibinfo {author} {\bibfnamefont
  {F.}~\bibnamefont {Mart{\'\i}n}}, \ and\ \bibinfo {author} {\bibfnamefont
  {A.~L.~V.}\ \bibnamefont {de~Parga}},\ }\href
  {https://arxiv.org/abs/2205.11383} {\bibfield  {journal} {\bibinfo  {journal}
  {arXiv:2205.11383}\ } (\bibinfo {year} {2022})}\BibitemShut {NoStop}%
\bibitem [{\citenamefont {Qiao}\ \emph {et~al.}(2017)\citenamefont {Qiao},
  \citenamefont {Li}, \citenamefont {Wang}, \citenamefont {Ruan}, \citenamefont
  {Ye}, \citenamefont {Cai}, \citenamefont {Hao}, \citenamefont {Yao},
  \citenamefont {Chen}, \citenamefont {Wu}, \citenamefont {Wang},\ and\
  \citenamefont {Liu}}]{QiaoLiu2017}%
  \BibitemOpen
  \bibfield  {author} {\bibinfo {author} {\bibfnamefont {S.}~\bibnamefont
  {Qiao}}, \bibinfo {author} {\bibfnamefont {X.}~\bibnamefont {Li}}, \bibinfo
  {author} {\bibfnamefont {N.}~\bibnamefont {Wang}}, \bibinfo {author}
  {\bibfnamefont {W.}~\bibnamefont {Ruan}}, \bibinfo {author} {\bibfnamefont
  {C.}~\bibnamefont {Ye}}, \bibinfo {author} {\bibfnamefont {P.}~\bibnamefont
  {Cai}}, \bibinfo {author} {\bibfnamefont {Z.}~\bibnamefont {Hao}}, \bibinfo
  {author} {\bibfnamefont {H.}~\bibnamefont {Yao}}, \bibinfo {author}
  {\bibfnamefont {X.}~\bibnamefont {Chen}}, \bibinfo {author} {\bibfnamefont
  {J.}~\bibnamefont {Wu}}, \bibinfo {author} {\bibfnamefont {Y.}~\bibnamefont
  {Wang}}, \ and\ \bibinfo {author} {\bibfnamefont {Z.}~\bibnamefont {Liu}},\
  }\href {\doibase 10.1103/PhysRevX.7.041054} {\bibfield  {journal} {\bibinfo
  {journal} {Phys. Rev. X}\ }\textbf {\bibinfo {volume} {7}},\ \bibinfo {pages}
  {041054} (\bibinfo {year} {2017})}\BibitemShut {NoStop}%
\bibitem [{\citenamefont {Chen}\ \emph {et~al.}(2020)\citenamefont {Chen},
  \citenamefont {Ruan}, \citenamefont {Wu}, \citenamefont {Tang}, \citenamefont
  {Ryu}, \citenamefont {Tsai}, \citenamefont {Lee}, \citenamefont {Kahn},
  \citenamefont {Liou}, \citenamefont {Jia} \emph {et~al.}}]{ChenCrommie2020}%
  \BibitemOpen
  \bibfield  {author} {\bibinfo {author} {\bibfnamefont {Y.}~\bibnamefont
  {Chen}}, \bibinfo {author} {\bibfnamefont {W.}~\bibnamefont {Ruan}}, \bibinfo
  {author} {\bibfnamefont {M.}~\bibnamefont {Wu}}, \bibinfo {author}
  {\bibfnamefont {S.}~\bibnamefont {Tang}}, \bibinfo {author} {\bibfnamefont
  {H.}~\bibnamefont {Ryu}}, \bibinfo {author} {\bibfnamefont {H.-Z.}\
  \bibnamefont {Tsai}}, \bibinfo {author} {\bibfnamefont {R.}~\bibnamefont
  {Lee}}, \bibinfo {author} {\bibfnamefont {S.}~\bibnamefont {Kahn}}, \bibinfo
  {author} {\bibfnamefont {F.}~\bibnamefont {Liou}}, \bibinfo {author}
  {\bibfnamefont {C.}~\bibnamefont {Jia}},  \emph {et~al.},\ }\href
  {https://www.nature.com/articles/s41567-019-0744-9} {\bibfield  {journal}
  {\bibinfo  {journal} {Nature Physics}\ }\textbf {\bibinfo {volume} {16}},\
  \bibinfo {pages} {218} (\bibinfo {year} {2020})}\BibitemShut {NoStop}%
\bibitem [{\citenamefont {Ruan}\ \emph {et~al.}(2021)\citenamefont {Ruan},
  \citenamefont {Chen}, \citenamefont {Tang}, \citenamefont {Hwang},
  \citenamefont {Tsai}, \citenamefont {Lee}, \citenamefont {Wu}, \citenamefont
  {Ryu}, \citenamefont {Kahn}, \citenamefont {Liou} \emph
  {et~al.}}]{RuanCrommie2021}%
  \BibitemOpen
  \bibfield  {author} {\bibinfo {author} {\bibfnamefont {W.}~\bibnamefont
  {Ruan}}, \bibinfo {author} {\bibfnamefont {Y.}~\bibnamefont {Chen}}, \bibinfo
  {author} {\bibfnamefont {S.}~\bibnamefont {Tang}}, \bibinfo {author}
  {\bibfnamefont {J.}~\bibnamefont {Hwang}}, \bibinfo {author} {\bibfnamefont
  {H.-Z.}\ \bibnamefont {Tsai}}, \bibinfo {author} {\bibfnamefont {R.~L.}\
  \bibnamefont {Lee}}, \bibinfo {author} {\bibfnamefont {M.}~\bibnamefont
  {Wu}}, \bibinfo {author} {\bibfnamefont {H.}~\bibnamefont {Ryu}}, \bibinfo
  {author} {\bibfnamefont {S.}~\bibnamefont {Kahn}}, \bibinfo {author}
  {\bibfnamefont {F.}~\bibnamefont {Liou}},  \emph {et~al.},\ }\href
  {https://doi.org/10.1038/s41567-021-01321-0} {\bibfield  {journal} {\bibinfo
  {journal} {Nature Physics}\ } (\bibinfo {year} {2021})}\BibitemShut {NoStop}%
\bibitem [{\citenamefont {Wan}\ \emph {et~al.}(2022)\citenamefont {Wan},
  \citenamefont {Harsh}, \citenamefont {Meninno}, \citenamefont {Dreher},
  \citenamefont {Sajan}, \citenamefont {Errea}, \citenamefont {de~Juan},\ and\
  \citenamefont {Ugeda}}]{WanUgeda2022}%
  \BibitemOpen
  \bibfield  {author} {\bibinfo {author} {\bibfnamefont {W.}~\bibnamefont
  {Wan}}, \bibinfo {author} {\bibfnamefont {R.}~\bibnamefont {Harsh}}, \bibinfo
  {author} {\bibfnamefont {A.}~\bibnamefont {Meninno}}, \bibinfo {author}
  {\bibfnamefont {P.}~\bibnamefont {Dreher}}, \bibinfo {author} {\bibfnamefont
  {S.}~\bibnamefont {Sajan}}, \bibinfo {author} {\bibfnamefont
  {I.}~\bibnamefont {Errea}}, \bibinfo {author} {\bibfnamefont
  {F.}~\bibnamefont {de~Juan}}, \ and\ \bibinfo {author} {\bibfnamefont
  {M.~M.}\ \bibnamefont {Ugeda}},\ }\href {https://arxiv.org/abs/2207.00096}
  {\bibfield  {journal} {\bibinfo  {journal} {arXiv:2207.00096}\ } (\bibinfo
  {year} {2022})}\BibitemShut {NoStop}%
\bibitem [{\citenamefont {Chen}\ \emph {et~al.}(2022)\citenamefont {Chen},
  \citenamefont {He}, \citenamefont {Ruan}, \citenamefont {Hwang},
  \citenamefont {Tang}, \citenamefont {Lee}, \citenamefont {Wu}, \citenamefont
  {Zhu}, \citenamefont {Zhang}, \citenamefont {Ryu} \emph
  {et~al.}}]{ChenCrommie2022}%
  \BibitemOpen
  \bibfield  {author} {\bibinfo {author} {\bibfnamefont {Y.}~\bibnamefont
  {Chen}}, \bibinfo {author} {\bibfnamefont {W.-Y.}\ \bibnamefont {He}},
  \bibinfo {author} {\bibfnamefont {W.}~\bibnamefont {Ruan}}, \bibinfo {author}
  {\bibfnamefont {J.}~\bibnamefont {Hwang}}, \bibinfo {author} {\bibfnamefont
  {S.}~\bibnamefont {Tang}}, \bibinfo {author} {\bibfnamefont {R.~L.}\
  \bibnamefont {Lee}}, \bibinfo {author} {\bibfnamefont {M.}~\bibnamefont
  {Wu}}, \bibinfo {author} {\bibfnamefont {T.}~\bibnamefont {Zhu}}, \bibinfo
  {author} {\bibfnamefont {C.}~\bibnamefont {Zhang}}, \bibinfo {author}
  {\bibfnamefont {H.}~\bibnamefont {Ryu}},  \emph {et~al.},\ }\href
  {https://www.nature.com/articles/s41567-022-01751-4} {\bibfield  {journal}
  {\bibinfo  {journal} {Nature Physics}\ }\textbf {\bibinfo {volume} {18}},\
  \bibinfo {pages} {1335} (\bibinfo {year} {2022})}\BibitemShut {NoStop}%
\bibitem [{\citenamefont {Liu}\ \emph {et~al.}(2021)\citenamefont {Liu},
  \citenamefont {Leveillee}, \citenamefont {Lu}, \citenamefont {Yu},
  \citenamefont {Kim}, \citenamefont {Tian}, \citenamefont {Shi}, \citenamefont
  {Lai}, \citenamefont {Zhang}, \citenamefont {Giustino} \emph
  {et~al.}}]{LiuShih2021}%
  \BibitemOpen
  \bibfield  {author} {\bibinfo {author} {\bibfnamefont {M.}~\bibnamefont
  {Liu}}, \bibinfo {author} {\bibfnamefont {J.}~\bibnamefont {Leveillee}},
  \bibinfo {author} {\bibfnamefont {S.}~\bibnamefont {Lu}}, \bibinfo {author}
  {\bibfnamefont {J.}~\bibnamefont {Yu}}, \bibinfo {author} {\bibfnamefont
  {H.}~\bibnamefont {Kim}}, \bibinfo {author} {\bibfnamefont {C.}~\bibnamefont
  {Tian}}, \bibinfo {author} {\bibfnamefont {Y.}~\bibnamefont {Shi}}, \bibinfo
  {author} {\bibfnamefont {K.}~\bibnamefont {Lai}}, \bibinfo {author}
  {\bibfnamefont {C.}~\bibnamefont {Zhang}}, \bibinfo {author} {\bibfnamefont
  {F.}~\bibnamefont {Giustino}},  \emph {et~al.},\ }\href
  {https://www.science.org/doi/full/10.1126/sciadv.abi6339} {\bibfield
  {journal} {\bibinfo  {journal} {Science advances}\ }\textbf {\bibinfo
  {volume} {7}},\ \bibinfo {pages} {eabi6339} (\bibinfo {year}
  {2021})}\BibitemShut {NoStop}%
\bibitem [{\citenamefont {Kratochvilova}\ \emph {et~al.}(2017)\citenamefont
  {Kratochvilova}, \citenamefont {Hillier}, \citenamefont {Wildes},
  \citenamefont {Wang}, \citenamefont {Cheong},\ and\ \citenamefont
  {Park}}]{KratochvilovaPark2017}%
  \BibitemOpen
  \bibfield  {author} {\bibinfo {author} {\bibfnamefont {M.}~\bibnamefont
  {Kratochvilova}}, \bibinfo {author} {\bibfnamefont {A.~D.}\ \bibnamefont
  {Hillier}}, \bibinfo {author} {\bibfnamefont {A.~R.}\ \bibnamefont {Wildes}},
  \bibinfo {author} {\bibfnamefont {L.}~\bibnamefont {Wang}}, \bibinfo {author}
  {\bibfnamefont {S.-W.}\ \bibnamefont {Cheong}}, \ and\ \bibinfo {author}
  {\bibfnamefont {J.-G.}\ \bibnamefont {Park}},\ }\href
  {https://www.nature.com/articles/s41535-017-0048-1} {\bibfield  {journal}
  {\bibinfo  {journal} {npj Quantum Materials}\ }\textbf {\bibinfo {volume}
  {2}},\ \bibinfo {pages} {1} (\bibinfo {year} {2017})}\BibitemShut {NoStop}%
\bibitem [{\citenamefont {Klanj{{s}}ek}\ \emph {et~al.}(2017)\citenamefont
  {Klanj{{s}}ek}, \citenamefont {Zorko}, \citenamefont {Mravlje}, \citenamefont
  {Jagli{{c}}i{\'c}}, \citenamefont {Biswas}, \citenamefont {Prelov{{s}}ek},
  \citenamefont {Mihailovic}, \citenamefont {Ar{{c}}on} \emph
  {et~al.}}]{Klanjvsek2017}%
  \BibitemOpen
  \bibfield  {author} {\bibinfo {author} {\bibfnamefont {M.}~\bibnamefont
  {Klanj{{s}}ek}}, \bibinfo {author} {\bibfnamefont {A.}~\bibnamefont {Zorko}},
  \bibinfo {author} {\bibfnamefont {J.}~\bibnamefont {Mravlje}}, \bibinfo
  {author} {\bibfnamefont {Z.}~\bibnamefont {Jagli{{c}}i{\'c}}}, \bibinfo
  {author} {\bibfnamefont {P.~K.}\ \bibnamefont {Biswas}}, \bibinfo {author}
  {\bibfnamefont {P.}~\bibnamefont {Prelov{{s}}ek}}, \bibinfo {author}
  {\bibfnamefont {D.}~\bibnamefont {Mihailovic}}, \bibinfo {author}
  {\bibfnamefont {D.}~\bibnamefont {Ar{{c}}on}},  \emph {et~al.},\ }\href
  {https://www.nature.com/articles/nphys4212} {\bibfield  {journal} {\bibinfo
  {journal} {Nature Physics}\ }\textbf {\bibinfo {volume} {13}},\ \bibinfo
  {pages} {1130} (\bibinfo {year} {2017})}\BibitemShut {NoStop}%
\bibitem [{\citenamefont {Law}\ and\ \citenamefont {Lee}(2017)}]{LawLee2017}%
  \BibitemOpen
  \bibfield  {author} {\bibinfo {author} {\bibfnamefont {K.~T.}\ \bibnamefont
  {Law}}\ and\ \bibinfo {author} {\bibfnamefont {P.~A.}\ \bibnamefont {Lee}},\
  }\href {https://www.pnas.org/content/114/27/6996.short} {\bibfield  {journal}
  {\bibinfo  {journal} {Proceedings of the National Academy of Sciences}\
  }\textbf {\bibinfo {volume} {114}},\ \bibinfo {pages} {6996} (\bibinfo {year}
  {2017})}\BibitemShut {NoStop}%
\bibitem [{\citenamefont {He}\ \emph {et~al.}(2018)\citenamefont {He},
  \citenamefont {Xu}, \citenamefont {Chen}, \citenamefont {Law},\ and\
  \citenamefont {Lee}}]{HeLee2018}%
  \BibitemOpen
  \bibfield  {author} {\bibinfo {author} {\bibfnamefont {W.-Y.}\ \bibnamefont
  {He}}, \bibinfo {author} {\bibfnamefont {X.~Y.}\ \bibnamefont {Xu}}, \bibinfo
  {author} {\bibfnamefont {G.}~\bibnamefont {Chen}}, \bibinfo {author}
  {\bibfnamefont {K.~T.}\ \bibnamefont {Law}}, \ and\ \bibinfo {author}
  {\bibfnamefont {P.~A.}\ \bibnamefont {Lee}},\ }\href {\doibase
  10.1103/PhysRevLett.121.046401} {\bibfield  {journal} {\bibinfo  {journal}
  {Phys. Rev. Lett.}\ }\textbf {\bibinfo {volume} {121}},\ \bibinfo {pages}
  {046401} (\bibinfo {year} {2018})}\BibitemShut {NoStop}%
\bibitem [{\citenamefont {Savary}\ and\ \citenamefont
  {Balents}(2016)}]{SavaryBalents2016}%
  \BibitemOpen
  \bibfield  {author} {\bibinfo {author} {\bibfnamefont {L.}~\bibnamefont
  {Savary}}\ and\ \bibinfo {author} {\bibfnamefont {L.}~\bibnamefont
  {Balents}},\ }\href
  {https://iopscience.iop.org/article/10.1088/0034-4885/80/1/016502/meta}
  {\bibfield  {journal} {\bibinfo  {journal} {Reports on Progress in Physics}\
  }\textbf {\bibinfo {volume} {80}},\ \bibinfo {pages} {016502} (\bibinfo
  {year} {2016})}\BibitemShut {NoStop}%
\bibitem [{\citenamefont {Zhou}\ \emph {et~al.}(2017)\citenamefont {Zhou},
  \citenamefont {Kanoda},\ and\ \citenamefont {Ng}}]{ZhouNg2017}%
  \BibitemOpen
  \bibfield  {author} {\bibinfo {author} {\bibfnamefont {Y.}~\bibnamefont
  {Zhou}}, \bibinfo {author} {\bibfnamefont {K.}~\bibnamefont {Kanoda}}, \ and\
  \bibinfo {author} {\bibfnamefont {T.-K.}\ \bibnamefont {Ng}},\ }\href
  {\doibase 10.1103/RevModPhys.89.025003} {\bibfield  {journal} {\bibinfo
  {journal} {Rev. Mod. Phys.}\ }\textbf {\bibinfo {volume} {89}},\ \bibinfo
  {pages} {025003} (\bibinfo {year} {2017})}\BibitemShut {NoStop}%
\bibitem [{\citenamefont {Ribak}\ \emph {et~al.}(2020)\citenamefont {Ribak},
  \citenamefont {Skiff}, \citenamefont {Mograbi}, \citenamefont {Rout},
  \citenamefont {Fischer}, \citenamefont {Ruhman}, \citenamefont {Chashka},
  \citenamefont {Dagan},\ and\ \citenamefont {Kanigel}}]{RibakKanigel2020}%
  \BibitemOpen
  \bibfield  {author} {\bibinfo {author} {\bibfnamefont {A.}~\bibnamefont
  {Ribak}}, \bibinfo {author} {\bibfnamefont {R.~M.}\ \bibnamefont {Skiff}},
  \bibinfo {author} {\bibfnamefont {M.}~\bibnamefont {Mograbi}}, \bibinfo
  {author} {\bibfnamefont {P.}~\bibnamefont {Rout}}, \bibinfo {author}
  {\bibfnamefont {M.}~\bibnamefont {Fischer}}, \bibinfo {author} {\bibfnamefont
  {J.}~\bibnamefont {Ruhman}}, \bibinfo {author} {\bibfnamefont
  {K.}~\bibnamefont {Chashka}}, \bibinfo {author} {\bibfnamefont
  {Y.}~\bibnamefont {Dagan}}, \ and\ \bibinfo {author} {\bibfnamefont
  {A.}~\bibnamefont {Kanigel}},\ }\href
  {https://advances.sciencemag.org/content/6/13/eaax9480.abstract} {\bibfield
  {journal} {\bibinfo  {journal} {Science advances}\ }\textbf {\bibinfo
  {volume} {6}},\ \bibinfo {pages} {eaax9480} (\bibinfo {year}
  {2020})}\BibitemShut {NoStop}%
\bibitem [{\citenamefont {Nayak}\ \emph {et~al.}(2021)\citenamefont {Nayak},
  \citenamefont {Steinbok}, \citenamefont {Roet}, \citenamefont {Koo},
  \citenamefont {Margalit}, \citenamefont {Feldman}, \citenamefont {Almoalem},
  \citenamefont {Kanigel}, \citenamefont {Fiete}, \citenamefont {Yan} \emph
  {et~al.}}]{NayakBeidenkopf2021}%
  \BibitemOpen
  \bibfield  {author} {\bibinfo {author} {\bibfnamefont {A.~K.}\ \bibnamefont
  {Nayak}}, \bibinfo {author} {\bibfnamefont {A.}~\bibnamefont {Steinbok}},
  \bibinfo {author} {\bibfnamefont {Y.}~\bibnamefont {Roet}}, \bibinfo {author}
  {\bibfnamefont {J.}~\bibnamefont {Koo}}, \bibinfo {author} {\bibfnamefont
  {G.}~\bibnamefont {Margalit}}, \bibinfo {author} {\bibfnamefont
  {I.}~\bibnamefont {Feldman}}, \bibinfo {author} {\bibfnamefont
  {A.}~\bibnamefont {Almoalem}}, \bibinfo {author} {\bibfnamefont
  {A.}~\bibnamefont {Kanigel}}, \bibinfo {author} {\bibfnamefont {G.~A.}\
  \bibnamefont {Fiete}}, \bibinfo {author} {\bibfnamefont {B.}~\bibnamefont
  {Yan}},  \emph {et~al.},\ }\href
  {https://www.nature.com/articles/s41567-021-01376-z} {\bibfield  {journal}
  {\bibinfo  {journal} {Nature physics}\ }\textbf {\bibinfo {volume} {17}},\
  \bibinfo {pages} {1413} (\bibinfo {year} {2021})}\BibitemShut {NoStop}%
\bibitem [{\citenamefont {Silber}\ \emph {et~al.}(2022)\citenamefont {Silber},
  \citenamefont {Mathimalar}, \citenamefont {Mangel}, \citenamefont {Green},
  \citenamefont {Avraham}, \citenamefont {Beidenkopf}, \citenamefont {Feldman},
  \citenamefont {Kanigel}, \citenamefont {Klein}, \citenamefont {Goldstein}
  \emph {et~al.}}]{SilberDagan2022}%
  \BibitemOpen
  \bibfield  {author} {\bibinfo {author} {\bibfnamefont {I.}~\bibnamefont
  {Silber}}, \bibinfo {author} {\bibfnamefont {S.}~\bibnamefont {Mathimalar}},
  \bibinfo {author} {\bibfnamefont {I.}~\bibnamefont {Mangel}}, \bibinfo
  {author} {\bibfnamefont {O.}~\bibnamefont {Green}}, \bibinfo {author}
  {\bibfnamefont {N.}~\bibnamefont {Avraham}}, \bibinfo {author} {\bibfnamefont
  {H.}~\bibnamefont {Beidenkopf}}, \bibinfo {author} {\bibfnamefont
  {I.}~\bibnamefont {Feldman}}, \bibinfo {author} {\bibfnamefont
  {A.}~\bibnamefont {Kanigel}}, \bibinfo {author} {\bibfnamefont
  {A.}~\bibnamefont {Klein}}, \bibinfo {author} {\bibfnamefont
  {M.}~\bibnamefont {Goldstein}},  \emph {et~al.},\ }\href
  {https://arxiv.org/abs/2208.14442} {\bibfield  {journal} {\bibinfo  {journal}
  {arXiv:2208.14442}\ } (\bibinfo {year} {2022})}\BibitemShut {NoStop}%
\bibitem [{\citenamefont {Almoalem}\ \emph {et~al.}(2022)\citenamefont
  {Almoalem}, \citenamefont {Feldman}, \citenamefont {Shlafman}, \citenamefont
  {Yaish}, \citenamefont {Fischer}, \citenamefont {Moshe}, \citenamefont
  {Ruhman},\ and\ \citenamefont {Kanigel}}]{AlmoalemKanigel2022}%
  \BibitemOpen
  \bibfield  {author} {\bibinfo {author} {\bibfnamefont {A.}~\bibnamefont
  {Almoalem}}, \bibinfo {author} {\bibfnamefont {I.}~\bibnamefont {Feldman}},
  \bibinfo {author} {\bibfnamefont {M.}~\bibnamefont {Shlafman}}, \bibinfo
  {author} {\bibfnamefont {Y.~E.}\ \bibnamefont {Yaish}}, \bibinfo {author}
  {\bibfnamefont {M.~H.}\ \bibnamefont {Fischer}}, \bibinfo {author}
  {\bibfnamefont {M.}~\bibnamefont {Moshe}}, \bibinfo {author} {\bibfnamefont
  {J.}~\bibnamefont {Ruhman}}, \ and\ \bibinfo {author} {\bibfnamefont
  {A.}~\bibnamefont {Kanigel}},\ }\href {https://arxiv.org/abs/2208.13798}
  {\bibfield  {journal} {\bibinfo  {journal} {arXiv:2208.13798}\ } (\bibinfo
  {year} {2022})}\BibitemShut {NoStop}%
\bibitem [{\citenamefont {Persky}\ \emph {et~al.}(2022)\citenamefont {Persky},
  \citenamefont {Bj{\o}rlig}, \citenamefont {Feldman}, \citenamefont
  {Almoalem}, \citenamefont {Altman}, \citenamefont {Berg}, \citenamefont
  {Kimchi}, \citenamefont {Ruhman}, \citenamefont {Kanigel},\ and\
  \citenamefont {Kalisky}}]{PerskyKalisky2022}%
  \BibitemOpen
  \bibfield  {author} {\bibinfo {author} {\bibfnamefont {E.}~\bibnamefont
  {Persky}}, \bibinfo {author} {\bibfnamefont {A.~V.}\ \bibnamefont
  {Bj{\o}rlig}}, \bibinfo {author} {\bibfnamefont {I.}~\bibnamefont {Feldman}},
  \bibinfo {author} {\bibfnamefont {A.}~\bibnamefont {Almoalem}}, \bibinfo
  {author} {\bibfnamefont {E.}~\bibnamefont {Altman}}, \bibinfo {author}
  {\bibfnamefont {E.}~\bibnamefont {Berg}}, \bibinfo {author} {\bibfnamefont
  {I.}~\bibnamefont {Kimchi}}, \bibinfo {author} {\bibfnamefont
  {J.}~\bibnamefont {Ruhman}}, \bibinfo {author} {\bibfnamefont
  {A.}~\bibnamefont {Kanigel}}, \ and\ \bibinfo {author} {\bibfnamefont
  {B.}~\bibnamefont {Kalisky}},\ }\href
  {https://www.nature.com/articles/s41586-022-04855-2} {\bibfield  {journal}
  {\bibinfo  {journal} {Nature}\ }\textbf {\bibinfo {volume} {607}},\ \bibinfo
  {pages} {692} (\bibinfo {year} {2022})}\BibitemShut {NoStop}%
\bibitem [{foo(oned)}]{footnoteCSL}%
  \BibitemOpen
  \href@noop {} {} (\bibinfo {year} {We note in passing that a scenario of a
  non-Abelian chiral spin liquid with gauge group SU(2) is also conceivable.
  The authors of the experimental paper~\cite{PerskyKalisky2022} did not
  elaborate in detail which CSL they envisioned})\BibitemShut {NoStop}%
\bibitem [{\citenamefont {Lin}(2022)}]{Lin2022}%
  \BibitemOpen
  \bibfield  {author} {\bibinfo {author} {\bibfnamefont {S.-Z.}\ \bibnamefont
  {Lin}},\ }\href {https://arxiv.org/abs/2210.06550} {\bibfield  {journal}
  {\bibinfo  {journal} {arXiv:2210.06550}\ } (\bibinfo {year}
  {2022})}\BibitemShut {NoStop}%
\bibitem [{\citenamefont {Chen}(2022)}]{Chen2023}%
  \BibitemOpen
  \bibfield  {author} {\bibinfo {author} {\bibfnamefont {G.}~\bibnamefont
  {Chen}},\ }\href {https://arxiv.org/abs/2208.03995} {\bibfield  {journal}
  {\bibinfo  {journal} {arXiv:2208.03995}\ } (\bibinfo {year}
  {2022})}\BibitemShut {NoStop}%
\bibitem [{\citenamefont {Fernandes}\ \emph {et~al.}(2019)\citenamefont
  {Fernandes}, \citenamefont {Orth},\ and\ \citenamefont
  {Schmalian}}]{FernandesSchmalian2019}%
  \BibitemOpen
  \bibfield  {author} {\bibinfo {author} {\bibfnamefont {R.~M.}\ \bibnamefont
  {Fernandes}}, \bibinfo {author} {\bibfnamefont {P.~P.}\ \bibnamefont {Orth}},
  \ and\ \bibinfo {author} {\bibfnamefont {J.}~\bibnamefont {Schmalian}},\
  }\href@noop {} {\bibfield  {journal} {\bibinfo  {journal} {Annual Review of
  Condensed Matter Physics}\ }\textbf {\bibinfo {volume} {10}},\ \bibinfo
  {pages} {133} (\bibinfo {year} {2019})}\BibitemShut {NoStop}%
\bibitem [{\citenamefont {Bojesen}\ \emph {et~al.}(2013)\citenamefont
  {Bojesen}, \citenamefont {Babaev},\ and\ \citenamefont
  {Sudb\o{}}}]{BojesenSudbo2013}%
  \BibitemOpen
  \bibfield  {author} {\bibinfo {author} {\bibfnamefont {T.~A.}\ \bibnamefont
  {Bojesen}}, \bibinfo {author} {\bibfnamefont {E.}~\bibnamefont {Babaev}}, \
  and\ \bibinfo {author} {\bibfnamefont {A.}~\bibnamefont {Sudb\o{}}},\ }\href
  {\doibase 10.1103/PhysRevB.88.220511} {\bibfield  {journal} {\bibinfo
  {journal} {Phys. Rev. B}\ }\textbf {\bibinfo {volume} {88}},\ \bibinfo
  {pages} {220511} (\bibinfo {year} {2013})}\BibitemShut {NoStop}%
\bibitem [{\citenamefont {Bojesen}\ \emph {et~al.}(2014)\citenamefont
  {Bojesen}, \citenamefont {Babaev},\ and\ \citenamefont
  {Sudb\o{}}}]{BojesenSudbo2014}%
  \BibitemOpen
  \bibfield  {author} {\bibinfo {author} {\bibfnamefont {T.~A.}\ \bibnamefont
  {Bojesen}}, \bibinfo {author} {\bibfnamefont {E.}~\bibnamefont {Babaev}}, \
  and\ \bibinfo {author} {\bibfnamefont {A.}~\bibnamefont {Sudb\o{}}},\ }\href
  {\doibase 10.1103/PhysRevB.89.104509} {\bibfield  {journal} {\bibinfo
  {journal} {Phys. Rev. B}\ }\textbf {\bibinfo {volume} {89}},\ \bibinfo
  {pages} {104509} (\bibinfo {year} {2014})}\BibitemShut {NoStop}%
\bibitem [{\citenamefont {Grinenko}\ \emph {et~al.}(2021)\citenamefont
  {Grinenko}, \citenamefont {Weston}, \citenamefont {Caglieris}, \citenamefont
  {Wuttke}, \citenamefont {Hess}, \citenamefont {Gottschall}, \citenamefont
  {Maccari}, \citenamefont {Gorbunov}, \citenamefont {Zherlitsyn},
  \citenamefont {Wosnitza}, \citenamefont {Rydh}, \citenamefont {Kihou},
  \citenamefont {Lee}, \citenamefont {Sarkar}, \citenamefont {Dengre},
  \citenamefont {Garaud}, \citenamefont {Charnukha}, \citenamefont {Hühne},
  \citenamefont {Nielsch}, \citenamefont {Büchner}, \citenamefont {Klauss},\
  and\ \citenamefont {Babaev}}]{GrinenkoBabaev2021}%
  \BibitemOpen
  \bibfield  {author} {\bibinfo {author} {\bibfnamefont {V.}~\bibnamefont
  {Grinenko}}, \bibinfo {author} {\bibfnamefont {D.}~\bibnamefont {Weston}},
  \bibinfo {author} {\bibfnamefont {F.}~\bibnamefont {Caglieris}}, \bibinfo
  {author} {\bibfnamefont {C.}~\bibnamefont {Wuttke}}, \bibinfo {author}
  {\bibfnamefont {C.}~\bibnamefont {Hess}}, \bibinfo {author} {\bibfnamefont
  {T.}~\bibnamefont {Gottschall}}, \bibinfo {author} {\bibfnamefont
  {I.}~\bibnamefont {Maccari}}, \bibinfo {author} {\bibfnamefont
  {D.}~\bibnamefont {Gorbunov}}, \bibinfo {author} {\bibfnamefont
  {S.}~\bibnamefont {Zherlitsyn}}, \bibinfo {author} {\bibfnamefont
  {J.}~\bibnamefont {Wosnitza}}, \bibinfo {author} {\bibfnamefont
  {A.}~\bibnamefont {Rydh}}, \bibinfo {author} {\bibfnamefont {K.}~\bibnamefont
  {Kihou}}, \bibinfo {author} {\bibfnamefont {C.-H.}\ \bibnamefont {Lee}},
  \bibinfo {author} {\bibfnamefont {R.}~\bibnamefont {Sarkar}}, \bibinfo
  {author} {\bibfnamefont {S.}~\bibnamefont {Dengre}}, \bibinfo {author}
  {\bibfnamefont {J.}~\bibnamefont {Garaud}}, \bibinfo {author} {\bibfnamefont
  {A.}~\bibnamefont {Charnukha}}, \bibinfo {author} {\bibfnamefont
  {R.}~\bibnamefont {Hühne}}, \bibinfo {author} {\bibfnamefont
  {K.}~\bibnamefont {Nielsch}}, \bibinfo {author} {\bibfnamefont
  {B.}~\bibnamefont {Büchner}}, \bibinfo {author} {\bibfnamefont {H.-H.}\
  \bibnamefont {Klauss}}, \ and\ \bibinfo {author} {\bibfnamefont
  {E.}~\bibnamefont {Babaev}},\ }\href {\doibase 10.1038/s41567-021-01350-9}
  {\bibfield  {journal} {\bibinfo  {journal} {Nature Physics}\ }\textbf
  {\bibinfo {volume} {17}},\ \bibinfo {pages} {1254} (\bibinfo {year}
  {2021})}\BibitemShut {NoStop}%
\bibitem [{\citenamefont {Crippa}\ \emph {et~al.}(2023)\citenamefont {Crippa},
  \citenamefont {Bae}, \citenamefont {Wunderlich}, \citenamefont {Mazin},
  \citenamefont {Yan}, \citenamefont {Sangiovanni}, \citenamefont {Wehling},\
  and\ \citenamefont {Valent{\'\i}}}]{CrippaValenti2023}%
  \BibitemOpen
  \bibfield  {author} {\bibinfo {author} {\bibfnamefont {L.}~\bibnamefont
  {Crippa}}, \bibinfo {author} {\bibfnamefont {H.}~\bibnamefont {Bae}},
  \bibinfo {author} {\bibfnamefont {P.}~\bibnamefont {Wunderlich}}, \bibinfo
  {author} {\bibfnamefont {I.~I.}\ \bibnamefont {Mazin}}, \bibinfo {author}
  {\bibfnamefont {B.}~\bibnamefont {Yan}}, \bibinfo {author} {\bibfnamefont
  {G.}~\bibnamefont {Sangiovanni}}, \bibinfo {author} {\bibfnamefont
  {T.}~\bibnamefont {Wehling}}, \ and\ \bibinfo {author} {\bibfnamefont
  {R.}~\bibnamefont {Valent{\'\i}}},\ }\href {https://arxiv.org/abs/2302.14072}
  {\bibfield  {journal} {\bibinfo  {journal} {arXiv:2302.14072}\ } (\bibinfo
  {year} {2023})}\BibitemShut {NoStop}%
\bibitem [{\citenamefont {Wen}\ \emph {et~al.}(2021)\citenamefont {Wen},
  \citenamefont {Gao}, \citenamefont {Xie}, \citenamefont {Zhang},
  \citenamefont {Kong}, \citenamefont {Wang}, \citenamefont {Jiang},
  \citenamefont {Luo}, \citenamefont {Li}, \citenamefont {Lu}, \citenamefont
  {Sun},\ and\ \citenamefont {Yan}}]{WenYan2021}%
  \BibitemOpen
  \bibfield  {author} {\bibinfo {author} {\bibfnamefont {C.}~\bibnamefont
  {Wen}}, \bibinfo {author} {\bibfnamefont {J.}~\bibnamefont {Gao}}, \bibinfo
  {author} {\bibfnamefont {Y.}~\bibnamefont {Xie}}, \bibinfo {author}
  {\bibfnamefont {Q.}~\bibnamefont {Zhang}}, \bibinfo {author} {\bibfnamefont
  {P.}~\bibnamefont {Kong}}, \bibinfo {author} {\bibfnamefont {J.}~\bibnamefont
  {Wang}}, \bibinfo {author} {\bibfnamefont {Y.}~\bibnamefont {Jiang}},
  \bibinfo {author} {\bibfnamefont {X.}~\bibnamefont {Luo}}, \bibinfo {author}
  {\bibfnamefont {J.}~\bibnamefont {Li}}, \bibinfo {author} {\bibfnamefont
  {W.}~\bibnamefont {Lu}}, \bibinfo {author} {\bibfnamefont {Y.-P.}\
  \bibnamefont {Sun}}, \ and\ \bibinfo {author} {\bibfnamefont
  {S.}~\bibnamefont {Yan}},\ }\href {\doibase 10.1103/PhysRevLett.126.256402}
  {\bibfield  {journal} {\bibinfo  {journal} {Phys. Rev. Lett.}\ }\textbf
  {\bibinfo {volume} {126}},\ \bibinfo {pages} {256402} (\bibinfo {year}
  {2021})}\BibitemShut {NoStop}%
\bibitem [{\citenamefont {Nayak}\ \emph {et~al.}(2023)\citenamefont {Nayak},
  \citenamefont {Steinbok}, \citenamefont {Roet}, \citenamefont {Koo},
  \citenamefont {Feldman}, \citenamefont {Almoalem}, \citenamefont {Kanigel},
  \citenamefont {Yan}, \citenamefont {Rosch}, \citenamefont {Avraham} \emph
  {et~al.}}]{NayakBeidenkopf2023}%
  \BibitemOpen
  \bibfield  {author} {\bibinfo {author} {\bibfnamefont {A.~K.}\ \bibnamefont
  {Nayak}}, \bibinfo {author} {\bibfnamefont {A.}~\bibnamefont {Steinbok}},
  \bibinfo {author} {\bibfnamefont {Y.}~\bibnamefont {Roet}}, \bibinfo {author}
  {\bibfnamefont {J.}~\bibnamefont {Koo}}, \bibinfo {author} {\bibfnamefont
  {I.}~\bibnamefont {Feldman}}, \bibinfo {author} {\bibfnamefont
  {A.}~\bibnamefont {Almoalem}}, \bibinfo {author} {\bibfnamefont
  {A.}~\bibnamefont {Kanigel}}, \bibinfo {author} {\bibfnamefont
  {B.}~\bibnamefont {Yan}}, \bibinfo {author} {\bibfnamefont {A.}~\bibnamefont
  {Rosch}}, \bibinfo {author} {\bibfnamefont {N.}~\bibnamefont {Avraham}},
  \emph {et~al.},\ }\href {https://arxiv.org/abs/2303.01447} {\bibfield
  {journal} {\bibinfo  {journal} {arXiv:2303.01447}\ } (\bibinfo {year}
  {2023})}\BibitemShut {NoStop}%
\bibitem [{foo(ible)}]{footnoteTopOrder}%
  \BibitemOpen
  \href@noop {} {} (\bibinfo {year} {We disagree with the statement of the
  author of \cite{Lin2022}, according to which condensation of a {K}ondo
  hybridization implies that the ``SU(2) gauge redundancy is broken down to
  U(1)''. Given that $V \sim \langle f^\dagger c \rangle$ transforms under the
  fundamental representation of the internal SU(2) gauge group, condensation
  will lead to Higgsing of all gauge fields and thus ultimately to confinement.
  Despite this disagreement, the overall proposal of Ref.~\cite{Lin2022} seems
  plausible})\BibitemShut {NoStop}%
\bibitem [{\citenamefont {Coleman}\ \emph {et~al.}(2005)\citenamefont
  {Coleman}, \citenamefont {Marston},\ and\ \citenamefont
  {Schofield}}]{ColemanSchofield2005}%
  \BibitemOpen
  \bibfield  {author} {\bibinfo {author} {\bibfnamefont {P.}~\bibnamefont
  {Coleman}}, \bibinfo {author} {\bibfnamefont {J.~B.}\ \bibnamefont
  {Marston}}, \ and\ \bibinfo {author} {\bibfnamefont {A.~J.}\ \bibnamefont
  {Schofield}},\ }\href {\doibase 10.1103/PhysRevB.72.245111} {\bibfield
  {journal} {\bibinfo  {journal} {Phys. Rev. B}\ }\textbf {\bibinfo {volume}
  {72}},\ \bibinfo {pages} {245111} (\bibinfo {year} {2005})}\BibitemShut
  {NoStop}%
\bibitem [{\citenamefont {Coleman}(2015)}]{ColemanBook}%
  \BibitemOpen
  \bibfield  {author} {\bibinfo {author} {\bibfnamefont {P.}~\bibnamefont
  {Coleman}},\ }\href@noop {} {\emph {\bibinfo {title} {Introduction to
  Many-Body Physics}}}\ (\bibinfo  {publisher} {Cambridge University Press},\
  \bibinfo {year} {2015})\BibitemShut {NoStop}%
\bibitem [{\citenamefont {Senthil}\ and\ \citenamefont
  {Fisher}(2001)}]{SenthilFisher2001}%
  \BibitemOpen
  \bibfield  {author} {\bibinfo {author} {\bibfnamefont {T.}~\bibnamefont
  {Senthil}}\ and\ \bibinfo {author} {\bibfnamefont {M.~P.~A.}\ \bibnamefont
  {Fisher}},\ }\href {\doibase 10.1103/PhysRevLett.86.292} {\bibfield
  {journal} {\bibinfo  {journal} {Phys. Rev. Lett.}\ }\textbf {\bibinfo
  {volume} {86}},\ \bibinfo {pages} {292} (\bibinfo {year} {2001})}\BibitemShut
  {NoStop}%
\bibitem [{\citenamefont {Cookmeyer}\ \emph {et~al.}(2021)\citenamefont
  {Cookmeyer}, \citenamefont {Motruk},\ and\ \citenamefont
  {Moore}}]{CookmeyerMoore2021}%
  \BibitemOpen
  \bibfield  {author} {\bibinfo {author} {\bibfnamefont {T.}~\bibnamefont
  {Cookmeyer}}, \bibinfo {author} {\bibfnamefont {J.}~\bibnamefont {Motruk}}, \
  and\ \bibinfo {author} {\bibfnamefont {J.~E.}\ \bibnamefont {Moore}},\ }\href
  {\doibase 10.1103/PhysRevLett.127.087201} {\bibfield  {journal} {\bibinfo
  {journal} {Phys. Rev. Lett.}\ }\textbf {\bibinfo {volume} {127}},\ \bibinfo
  {pages} {087201} (\bibinfo {year} {2021})}\BibitemShut {NoStop}%
\bibitem [{\citenamefont {Motrunich}(2006)}]{Motrunich2006}%
  \BibitemOpen
  \bibfield  {author} {\bibinfo {author} {\bibfnamefont {O.~I.}\ \bibnamefont
  {Motrunich}},\ }\href {\doibase 10.1103/PhysRevB.73.155115} {\bibfield
  {journal} {\bibinfo  {journal} {Phys. Rev. B}\ }\textbf {\bibinfo {volume}
  {73}},\ \bibinfo {pages} {155115} (\bibinfo {year} {2006})}\BibitemShut
  {NoStop}%
\bibitem [{\citenamefont {K\"onig}\ \emph {et~al.}(2021)\citenamefont
  {K\"onig}, \citenamefont {Coleman},\ and\ \citenamefont
  {Komijani}}]{KoenigKomijani2021}%
  \BibitemOpen
  \bibfield  {author} {\bibinfo {author} {\bibfnamefont {E.~J.}\ \bibnamefont
  {K\"onig}}, \bibinfo {author} {\bibfnamefont {P.}~\bibnamefont {Coleman}}, \
  and\ \bibinfo {author} {\bibfnamefont {Y.}~\bibnamefont {Komijani}},\ }\href
  {\doibase 10.1103/PhysRevB.104.115103} {\bibfield  {journal} {\bibinfo
  {journal} {Phys. Rev. B}\ }\textbf {\bibinfo {volume} {104}},\ \bibinfo
  {pages} {115103} (\bibinfo {year} {2021})}\BibitemShut {NoStop}%
\bibitem [{\citenamefont {Abrikosov}(1957)}]{Abrikosov1957}%
  \BibitemOpen
  \bibfield  {author} {\bibinfo {author} {\bibfnamefont {A.~A.}\ \bibnamefont
  {Abrikosov}},\ }\href@noop {} {\bibfield  {journal} {\bibinfo  {journal}
  {Soviet Physics-JETP}\ }\textbf {\bibinfo {volume} {5}},\ \bibinfo {pages}
  {1174} (\bibinfo {year} {1957})}\BibitemShut {NoStop}%
\bibitem [{\citenamefont {Rjabinin}\ and\ \citenamefont
  {Shubnikow}(1935)}]{RyabininShibnikov1935}%
  \BibitemOpen
  \bibfield  {author} {\bibinfo {author} {\bibfnamefont {J.~N.}\ \bibnamefont
  {Rjabinin}}\ and\ \bibinfo {author} {\bibfnamefont {L.}~\bibnamefont
  {Shubnikow}},\ }\href@noop {} {\bibfield  {journal} {\bibinfo  {journal}
  {Nature}\ }\textbf {\bibinfo {volume} {135}},\ \bibinfo {pages} {581}
  (\bibinfo {year} {1935})}\BibitemShut {NoStop}%
\bibitem [{\citenamefont {Read}\ and\ \citenamefont
  {Newns}(1983)}]{ReadNewns1983}%
  \BibitemOpen
  \bibfield  {author} {\bibinfo {author} {\bibfnamefont {N.}~\bibnamefont
  {Read}}\ and\ \bibinfo {author} {\bibfnamefont {D.}~\bibnamefont {Newns}},\
  }\href@noop {} {\bibfield  {journal} {\bibinfo  {journal} {Journal of Physics
  C: Solid State Physics}\ }\textbf {\bibinfo {volume} {16}},\ \bibinfo {pages}
  {3273} (\bibinfo {year} {1983})}\BibitemShut {NoStop}%
\bibitem [{\citenamefont {Coleman}(1983)}]{Coleman1983}%
  \BibitemOpen
  \bibfield  {author} {\bibinfo {author} {\bibfnamefont {P.}~\bibnamefont
  {Coleman}},\ }\href {\doibase 10.1103/PhysRevB.28.5255} {\bibfield  {journal}
  {\bibinfo  {journal} {Phys. Rev. B}\ }\textbf {\bibinfo {volume} {28}},\
  \bibinfo {pages} {5255} (\bibinfo {year} {1983})}\BibitemShut {NoStop}%
\bibitem [{\citenamefont {Barnes}(1976)}]{Barnes1976}%
  \BibitemOpen
  \bibfield  {author} {\bibinfo {author} {\bibfnamefont {S.}~\bibnamefont
  {Barnes}},\ }\href@noop {} {\bibfield  {journal} {\bibinfo  {journal}
  {Journal of Physics F: Metal Physics}\ }\textbf {\bibinfo {volume} {6}},\
  \bibinfo {pages} {1375} (\bibinfo {year} {1976})}\BibitemShut {NoStop}%
\bibitem [{\citenamefont {Coleman}(1984)}]{Coleman1984}%
  \BibitemOpen
  \bibfield  {author} {\bibinfo {author} {\bibfnamefont {P.}~\bibnamefont
  {Coleman}},\ }\href {\doibase 10.1103/PhysRevB.29.3035} {\bibfield  {journal}
  {\bibinfo  {journal} {Phys. Rev. B}\ }\textbf {\bibinfo {volume} {29}},\
  \bibinfo {pages} {3035} (\bibinfo {year} {1984})}\BibitemShut {NoStop}%
\bibitem [{\citenamefont {Senthil}\ \emph {et~al.}(2003)\citenamefont
  {Senthil}, \citenamefont {Sachdev},\ and\ \citenamefont
  {Vojta}}]{SenthilVojta2003}%
  \BibitemOpen
  \bibfield  {author} {\bibinfo {author} {\bibfnamefont {T.}~\bibnamefont
  {Senthil}}, \bibinfo {author} {\bibfnamefont {S.}~\bibnamefont {Sachdev}}, \
  and\ \bibinfo {author} {\bibfnamefont {M.}~\bibnamefont {Vojta}},\ }\href
  {\doibase 10.1103/PhysRevLett.90.216403} {\bibfield  {journal} {\bibinfo
  {journal} {Phys. Rev. Lett.}\ }\textbf {\bibinfo {volume} {90}},\ \bibinfo
  {pages} {216403} (\bibinfo {year} {2003})}\BibitemShut {NoStop}%
\bibitem [{\citenamefont {Senthil}\ \emph {et~al.}(2004)\citenamefont
  {Senthil}, \citenamefont {Vojta},\ and\ \citenamefont
  {Sachdev}}]{SenthilSachdev2004}%
  \BibitemOpen
  \bibfield  {author} {\bibinfo {author} {\bibfnamefont {T.}~\bibnamefont
  {Senthil}}, \bibinfo {author} {\bibfnamefont {M.}~\bibnamefont {Vojta}}, \
  and\ \bibinfo {author} {\bibfnamefont {S.}~\bibnamefont {Sachdev}},\ }\href
  {\doibase 10.1103/PhysRevB.69.035111} {\bibfield  {journal} {\bibinfo
  {journal} {Phys. Rev. B}\ }\textbf {\bibinfo {volume} {69}},\ \bibinfo
  {pages} {035111} (\bibinfo {year} {2004})}\BibitemShut {NoStop}%
\bibitem [{\citenamefont {Wen}\ and\ \citenamefont {Zee}(1989)}]{WenZee1989}%
  \BibitemOpen
  \bibfield  {author} {\bibinfo {author} {\bibfnamefont {X.~G.}\ \bibnamefont
  {Wen}}\ and\ \bibinfo {author} {\bibfnamefont {A.}~\bibnamefont {Zee}},\
  }\href {\doibase 10.1103/PhysRevLett.62.2873} {\bibfield  {journal} {\bibinfo
   {journal} {Phys. Rev. Lett.}\ }\textbf {\bibinfo {volume} {62}},\ \bibinfo
  {pages} {2873} (\bibinfo {year} {1989})}\BibitemShut {NoStop}%
\bibitem [{\citenamefont {Sachdev}(1992)}]{Sachdev1992}%
  \BibitemOpen
  \bibfield  {author} {\bibinfo {author} {\bibfnamefont {S.}~\bibnamefont
  {Sachdev}},\ }\href {\doibase 10.1103/PhysRevB.45.389} {\bibfield  {journal}
  {\bibinfo  {journal} {Phys. Rev. B}\ }\textbf {\bibinfo {volume} {45}},\
  \bibinfo {pages} {389} (\bibinfo {year} {1992})}\BibitemShut {NoStop}%
\bibitem [{\citenamefont {Nagaosa}\ and\ \citenamefont
  {Lee}(1992)}]{NagaosaLee1992}%
  \BibitemOpen
  \bibfield  {author} {\bibinfo {author} {\bibfnamefont {N.}~\bibnamefont
  {Nagaosa}}\ and\ \bibinfo {author} {\bibfnamefont {P.~A.}\ \bibnamefont
  {Lee}},\ }\href {\doibase 10.1103/PhysRevB.45.966} {\bibfield  {journal}
  {\bibinfo  {journal} {Phys. Rev. B}\ }\textbf {\bibinfo {volume} {45}},\
  \bibinfo {pages} {966} (\bibinfo {year} {1992})}\BibitemShut {NoStop}%
\bibitem [{\citenamefont {Tsvelik}\ and\ \citenamefont
  {Coleman}(2022)}]{TsvelikColeman2022}%
  \BibitemOpen
  \bibfield  {author} {\bibinfo {author} {\bibfnamefont {A.~M.}\ \bibnamefont
  {Tsvelik}}\ and\ \bibinfo {author} {\bibfnamefont {P.}~\bibnamefont
  {Coleman}},\ }\href {\doibase 10.1103/PhysRevB.106.125144} {\bibfield
  {journal} {\bibinfo  {journal} {Phys. Rev. B}\ }\textbf {\bibinfo {volume}
  {106}},\ \bibinfo {pages} {125144} (\bibinfo {year} {2022})}\BibitemShut
  {NoStop}%
\bibitem [{\citenamefont {Wugalter}\ \emph {et~al.}(2020)\citenamefont
  {Wugalter}, \citenamefont {Komijani},\ and\ \citenamefont
  {Coleman}}]{WugalterColeman2020}%
  \BibitemOpen
  \bibfield  {author} {\bibinfo {author} {\bibfnamefont {A.}~\bibnamefont
  {Wugalter}}, \bibinfo {author} {\bibfnamefont {Y.}~\bibnamefont {Komijani}},
  \ and\ \bibinfo {author} {\bibfnamefont {P.}~\bibnamefont {Coleman}},\ }\href
  {\doibase 10.1103/PhysRevB.101.075133} {\bibfield  {journal} {\bibinfo
  {journal} {Phys. Rev. B}\ }\textbf {\bibinfo {volume} {101}},\ \bibinfo
  {pages} {075133} (\bibinfo {year} {2020})}\BibitemShut {NoStop}%
\bibitem [{\citenamefont {Kornjaca}\ \emph {et~al.}(2021)\citenamefont
  {Kornjaca}, \citenamefont {Quito},\ and\ \citenamefont
  {Flint}}]{KornjavcaFlint2021}%
  \BibitemOpen
  \bibfield  {author} {\bibinfo {author} {\bibfnamefont {M.}~\bibnamefont
  {Kornjaca}}, \bibinfo {author} {\bibfnamefont {V.~L.}\ \bibnamefont {Quito}},
  \ and\ \bibinfo {author} {\bibfnamefont {R.}~\bibnamefont {Flint}},\ }\href
  {https://arxiv.org/abs/2104.11173} {\bibfield  {journal} {\bibinfo  {journal}
  {arXiv:2104.11173}\ } (\bibinfo {year} {2021})}\BibitemShut {NoStop}%
\bibitem [{\citenamefont {Ge}\ and\ \citenamefont
  {Komijani}(2022)}]{GeKomijani2022}%
  \BibitemOpen
  \bibfield  {author} {\bibinfo {author} {\bibfnamefont {Y.}~\bibnamefont
  {Ge}}\ and\ \bibinfo {author} {\bibfnamefont {Y.}~\bibnamefont {Komijani}},\
  }\href {\doibase 10.1103/PhysRevLett.129.077202} {\bibfield  {journal}
  {\bibinfo  {journal} {Phys. Rev. Lett.}\ }\textbf {\bibinfo {volume} {129}},\
  \bibinfo {pages} {077202} (\bibinfo {year} {2022})}\BibitemShut {NoStop}%
\bibitem [{\citenamefont {Saremi}\ \emph {et~al.}(2009)\citenamefont {Saremi},
  \citenamefont {Lee},\ and\ \citenamefont {Senthil}}]{SaremiSenthil2009}%
  \BibitemOpen
  \bibfield  {author} {\bibinfo {author} {\bibfnamefont {S.}~\bibnamefont
  {Saremi}}, \bibinfo {author} {\bibfnamefont {P.~A.}\ \bibnamefont {Lee}}, \
  and\ \bibinfo {author} {\bibfnamefont {T.}~\bibnamefont {Senthil}},\ }\href
  {https://arxiv.org/abs/0903.4195} {\bibfield  {journal} {\bibinfo  {journal}
  {arXiv:0903.4195}\ } (\bibinfo {year} {2009})}\BibitemShut {NoStop}%
\bibitem [{\citenamefont {Saremi}\ \emph {et~al.}(2011)\citenamefont {Saremi},
  \citenamefont {Lee},\ and\ \citenamefont {Senthil}}]{SaremiSenthil2011}%
  \BibitemOpen
  \bibfield  {author} {\bibinfo {author} {\bibfnamefont {S.}~\bibnamefont
  {Saremi}}, \bibinfo {author} {\bibfnamefont {P.~A.}\ \bibnamefont {Lee}}, \
  and\ \bibinfo {author} {\bibfnamefont {T.}~\bibnamefont {Senthil}},\ }\href
  {\doibase 10.1103/PhysRevB.83.125120} {\bibfield  {journal} {\bibinfo
  {journal} {Phys. Rev. B}\ }\textbf {\bibinfo {volume} {83}},\ \bibinfo
  {pages} {125120} (\bibinfo {year} {2011})}\BibitemShut {NoStop}%
\bibitem [{\citenamefont {Iimura}\ \emph {et~al.}(2020)\citenamefont {Iimura},
  \citenamefont {Hirayama},\ and\ \citenamefont {Hoshino}}]{IimuraHoshino2020}%
  \BibitemOpen
  \bibfield  {author} {\bibinfo {author} {\bibfnamefont {S.}~\bibnamefont
  {Iimura}}, \bibinfo {author} {\bibfnamefont {M.}~\bibnamefont {Hirayama}}, \
  and\ \bibinfo {author} {\bibfnamefont {S.}~\bibnamefont {Hoshino}},\ }\href
  {\doibase 10.1103/PhysRevB.102.064505} {\bibfield  {journal} {\bibinfo
  {journal} {Phys. Rev. B}\ }\textbf {\bibinfo {volume} {102}},\ \bibinfo
  {pages} {064505} (\bibinfo {year} {2020})}\BibitemShut {NoStop}%
\bibitem [{foo(sued)}]{footnoteVortex}%
  \BibitemOpen
  \href@noop {} {} (\bibinfo {year} {The possibility of vortex solutions in the
  Kondo hybridization was also mentioned in S.Z. Lin's~\cite{Lin2022} paper on
  the magnetic memory effect, but not further pursued})\BibitemShut {NoStop}%
\bibitem [{\citenamefont {Ramires}\ \emph {et~al.}(2012)\citenamefont
  {Ramires}, \citenamefont {Coleman}, \citenamefont {Nevidomskyy},\ and\
  \citenamefont {Tsvelik}}]{RamiresTsvelik2012}%
  \BibitemOpen
  \bibfield  {author} {\bibinfo {author} {\bibfnamefont {A.}~\bibnamefont
  {Ramires}}, \bibinfo {author} {\bibfnamefont {P.}~\bibnamefont {Coleman}},
  \bibinfo {author} {\bibfnamefont {A.~H.}\ \bibnamefont {Nevidomskyy}}, \ and\
  \bibinfo {author} {\bibfnamefont {A.~M.}\ \bibnamefont {Tsvelik}},\ }\href
  {\doibase 10.1103/PhysRevLett.109.176404} {\bibfield  {journal} {\bibinfo
  {journal} {Phys. Rev. Lett.}\ }\textbf {\bibinfo {volume} {109}},\ \bibinfo
  {pages} {176404} (\bibinfo {year} {2012})}\BibitemShut {NoStop}%
\bibitem [{\citenamefont {Elitzur}(1975)}]{Elitzur1975}%
  \BibitemOpen
  \bibfield  {author} {\bibinfo {author} {\bibfnamefont {S.}~\bibnamefont
  {Elitzur}},\ }\href {\doibase 10.1103/PhysRevD.12.3978} {\bibfield  {journal}
  {\bibinfo  {journal} {Phys. Rev. D}\ }\textbf {\bibinfo {volume} {12}},\
  \bibinfo {pages} {3978} (\bibinfo {year} {1975})}\BibitemShut {NoStop}%
\bibitem [{Sup()}]{SuppMat}%
  \BibitemOpen
  \href@noop {} {}\bibinfo {note} {See supplementary materials for a discussion
  of vortex solutions in the heavy Fermi liquid, for a microscopic derivation
  of Ginzburg-Landau parameters, for a discussion of subgap states trapped in
  the $V$-vortices as well as
  Refs.~\cite{LifshitzPitavskii2013,AuerbachBook}}\BibitemShut {NoStop}%
\bibitem [{\citenamefont {P\'epin}(2008)}]{Pepin2008}%
  \BibitemOpen
  \bibfield  {author} {\bibinfo {author} {\bibfnamefont {C.}~\bibnamefont
  {P\'epin}},\ }\href {\doibase 10.1103/PhysRevB.77.245129} {\bibfield
  {journal} {\bibinfo  {journal} {Phys. Rev. B}\ }\textbf {\bibinfo {volume}
  {77}},\ \bibinfo {pages} {245129} (\bibinfo {year} {2008})}\BibitemShut
  {NoStop}%
\bibitem [{\citenamefont {C\^onsoli}\ and\ \citenamefont
  {Vojta}(2022)}]{ConsoliVojta2022}%
  \BibitemOpen
  \bibfield  {author} {\bibinfo {author} {\bibfnamefont {P.~M.}\ \bibnamefont
  {C\^onsoli}}\ and\ \bibinfo {author} {\bibfnamefont {M.}~\bibnamefont
  {Vojta}},\ }\href {\doibase 10.1103/PhysRevB.106.235127} {\bibfield
  {journal} {\bibinfo  {journal} {Phys. Rev. B}\ }\textbf {\bibinfo {volume}
  {106}},\ \bibinfo {pages} {235127} (\bibinfo {year} {2022})}\BibitemShut
  {NoStop}%
\bibitem [{\citenamefont {Maltseva}\ \emph {et~al.}(2009)\citenamefont
  {Maltseva}, \citenamefont {Dzero},\ and\ \citenamefont
  {Coleman}}]{MaltsevaColeman2009}%
  \BibitemOpen
  \bibfield  {author} {\bibinfo {author} {\bibfnamefont {M.}~\bibnamefont
  {Maltseva}}, \bibinfo {author} {\bibfnamefont {M.}~\bibnamefont {Dzero}}, \
  and\ \bibinfo {author} {\bibfnamefont {P.}~\bibnamefont {Coleman}},\ }\href
  {\doibase 10.1103/PhysRevLett.103.206402} {\bibfield  {journal} {\bibinfo
  {journal} {Phys. Rev. Lett.}\ }\textbf {\bibinfo {volume} {103}},\ \bibinfo
  {pages} {206402} (\bibinfo {year} {2009})}\BibitemShut {NoStop}%
\bibitem [{\citenamefont {Caroli}\ \emph {et~al.}(1964)\citenamefont {Caroli},
  \citenamefont {De~Gennes},\ and\ \citenamefont
  {Matricon}}]{CaroliMatricon1964}%
  \BibitemOpen
  \bibfield  {author} {\bibinfo {author} {\bibfnamefont {C.}~\bibnamefont
  {Caroli}}, \bibinfo {author} {\bibfnamefont {P.}~\bibnamefont {De~Gennes}}, \
  and\ \bibinfo {author} {\bibfnamefont {J.}~\bibnamefont {Matricon}},\
  }\href@noop {} {\bibfield  {journal} {\bibinfo  {journal} {Physics Letters}\
  }\textbf {\bibinfo {volume} {9}},\ \bibinfo {pages} {307} (\bibinfo {year}
  {1964})}\BibitemShut {NoStop}%
\bibitem [{\citenamefont {Lifshitz}\ and\ \citenamefont
  {Pitaevskii}(2013)}]{LifshitzPitavskii2013}%
  \BibitemOpen
  \bibfield  {author} {\bibinfo {author} {\bibfnamefont {E.}~\bibnamefont
  {Lifshitz}}\ and\ \bibinfo {author} {\bibfnamefont {L.}~\bibnamefont
  {Pitaevskii}},\ }\href@noop {} {\emph {\bibinfo {title} {Statistical Physics:
  Theory of the Condensed State}}},\ \bibinfo {series} {Course of Theoretical
  Physics}\ No.\ \bibinfo {number} {v. 9}\ (\bibinfo  {publisher} {Elsevier
  Science},\ \bibinfo {year} {2013})\BibitemShut {NoStop}%
\bibitem [{\citenamefont {Auerbach}(1998)}]{AuerbachBook}%
  \BibitemOpen
  \bibfield  {author} {\bibinfo {author} {\bibfnamefont {A.}~\bibnamefont
  {Auerbach}},\ }\href@noop {} {\emph {\bibinfo {title} {Interacting electrons
  and quantum magnetism}}}\ (\bibinfo  {publisher} {Springer Science \&
  Business Media},\ \bibinfo {year} {1998})\BibitemShut {NoStop}%
\end{thebibliography}%

\clearpage

\setcounter{equation}{0}
\setcounter{figure}{0}
\setcounter{section}{0}
\setcounter{table}{0}
\setcounter{page}{1}
\makeatletter
\renewcommand{\theequation}{S\arabic{equation}}
\renewcommand{\thesection}{S\arabic{section}}
\renewcommand{\thefigure}{S\arabic{figure}}
\renewcommand{\thepage}{S\arabic{page}}

\begin{widetext}
\begin{center}
Supplementary materials on \\
\textbf{"Type-II heavy Fermi liquids and the magnetic memory of 4Hb-TaS$_2$"}\\
{Elio J. K\"{o}nig}\\
\textit{Max-Planck-Institut f\"{u}r Festk\"{o}rperforschung, Heisenbergstra{\ss}e 1, 70569 Stuttgart, Germany}
\end{center}

These supplemental materials contain technical details regarding the vortex solutions of type-II heavy Fermi liquids, Sec.~\ref{sec:Vortices}, details about the derivation of the Ginzburg-Landau functional from microscopics, Sec.~\ref{sec:MicroscopicsSM}, as well as the solution of fermionic subgap states in vortices of $V$, Sec.~\ref{sec:CdM}. Citation numbers refer to the bibliography of the main text.
\end{widetext}

\section{Phenomenology of type-II heavy Fermi liquids}
\label{sec:Vortices}

In this supplement, we derive the energetic cost of vortices in $V$ and $\Delta$ in the presence of applied fields $\v H, \v h$. This will demonstrate the stability of the type-II heavy Fermi liquid on phenomenological grounds (as long as $\lambda_K$ exceeds $\xi_K$). Many of the calculations in this section are generalizations of textbook calculations for type-II superconductors~\cite{LifshitzPitavskii2013}.

\subsection{Ginzburg-Landau functional}
We briefly comment on the Ginzburg-Landau functional, Eq.~\eqref{eq:GL} of the main text. In the given units, magnetic fields have dimenstion $[B]=[b] = \sqrt{E/L^3}$ and $[e] = [e_*] = \sqrt{EL}^{-1}$, $[\alpha_{\rm SC}]=[\alpha_{\rm K}] = 1/L^3E $, $[m_{\rm SC}]=[m_{\rm K}] = EL $ , where $E$ ($L$) denote energy (length) dimensions. In particular, in these units and keeping $\hbar = 1$
\begin{equation}
    e^2 = \alpha_{\rm QED}/c = 1/(137 c),
\end{equation}
where $c$ is the speed of light.

\subsection{Upper critical fields}

The upper critical field corresponds to the shift in $T_c, T_K$ due to the zero point-motion of the states in the lowest Landau level. We solve the linearized mean-field equations
\begin{subequations}
\begin{align}
\left [\frac{(- i \boldsymbol \nabla - 2 e \mathbf A)^2}{2m_{\rm SC}} - \vert \alpha \vert \right ] \vec \Delta &= 0,\\
\left [\frac{(- i \boldsymbol \nabla -  e \mathbf A + e_*\mathbf a)^2}{2m_{\rm K}} - \vert \alpha_K \vert \right ] V &= 0,
\end{align}
\end{subequations}
Using the standard expression $\omega_c = q H/m c$ we obtain for the lowest Landau level
\begin{subequations}
\begin{align}
\left [\frac{e H}{m_{\rm SC}} - \vert \alpha \vert \right ] \vec \Delta &= 0,\\
\left [\frac{\vert e H - e_* h \vert}{2m_{\rm K}} - \vert \alpha_K \vert \right ] V &= 0.
\end{align}
\end{subequations}

Thus we find the following following two conditions for the upper critical fields
\begin{subequations}
\begin{align}
H - H_{\xi} &= 0,\\
\vert \frac{e}{e_*} H  - h\vert - 2h_{\xi_K} &= 0,
\end{align}
\label{eq:UpperCritFields}
\end{subequations}
where we introduced $H_\xi = 1/(2 e \xi^2)$, $h_{\xi_K} = 1/(2 e_* \xi_K^2)$. Note that the upper critical field of $h$ is not merely given by $h_{\xi_K}$ (Fig~\ref{fig:MagField} a) of the main text). 

\subsection{Field configurations about a vortex solution}

The Ginzburg-Landau functional, Eq.~\eqref{eq:GL} of the main text, implies below $T_c$ and $T_K$, respectively, that

\begin{subequations}
\begin{align}
    f_{\rm SC} &= \underbrace{\frac{\vert \vec \Delta \vert^2 (2e)^2}{2m_{\rm SC}}}_{\equiv \frac{1}{8 \pi \lambda^2}} (\boldsymbol{\nabla} \theta \frac{\Phi_0}{2\pi}-  \mathbf A)^2, \\
     f_{\rm K} & = \underbrace{\frac{\vert V\vert^2 (2e_*)^2}{2m_{\rm K}}}_{\equiv \frac{1}{8\pi \lambda_K^2}} (\frac{\Phi_K}{2\pi}\boldsymbol{\nabla} \theta_{\rm K} -  \frac{e}{2e_*} \mathbf A + \frac{1}{2}\mathbf a)^2.
\end{align}
\end{subequations}

We introduced the phase of $\vec \Delta$ ($V$) as $\theta$ ($\theta_K$), the penetration lengths defined in the main text and the flux quanta $\Phi_0 = \pi/e$ ($\Phi_K = \pi/e_*$).

We are interested in the result to zeroth order in external fields $\mathbf H, \mathbf h$. From the variation of the free energy with respect to $\mathbf A, \mathbf a$ we obtain

\begin{widetext}
\begin{subequations}
\begin{align}
    0 & = \frac{1}{4\pi \lambda^2} (\mathbf A - \frac{\Phi_0}{2\pi} \nabla \theta) + \frac{1}{8\pi \lambda_K^2} \frac{e}{e_*}(\frac{e}{2 e_*}\mathbf A - \frac{1}{2 }\mathbf a - \frac{\Phi_K}{2\pi} \nabla \theta_K) - \frac{1}{4\pi} (\Delta - \mathbf \nabla \mathbf \nabla^T) \mathbf A,\\
    0 & =  \frac{1}{8\pi \lambda_K^2} (\frac{1}{2 }\mathbf a - \frac{e}{2 e_*}\mathbf A + \frac{\Phi_K}{2\pi} \nabla \theta_K) - \frac{1}{4\pi} (\Delta - \mathbf \nabla \mathbf \nabla^T) \mathbf a.
\end{align}
\label{eq:MF}
\end{subequations}
We take the rotation of these equations and consider (potential) vortex solutions
\begin{align}
    \mathbf \nabla \times     \mathbf \nabla \theta  = \zeta 2\pi \delta(x)\delta(y) \hat e_z, \quad
    \mathbf \nabla \times     \mathbf \nabla \theta_K  = \zeta_K 2\pi \delta(x)\delta(y) \hat e_z,
\end{align}
where $\zeta, \zeta_K \in \mathbb Z$ is a priori not fixed.
We obtain for $\mathbf B = B \hat e_z$, $\mathbf b = b \hat e_z$

\begin{subequations}
\begin{align}
  \left (\frac{1}{\lambda^2} + \frac{e^2}{4 e_*^2 \lambda_K^2} - \Delta\right) B - \frac{e}{4 e_* \lambda_K^2} b&=  \left (\zeta \frac{\Phi_0}{\lambda^2}  
 + \zeta_K \frac{\Phi_K}{\lambda_K^2} \frac{e}{2e_*}\right)\delta(x) \delta(y) , \\
   \left (\frac{1}{4  \lambda_K^2} - \Delta\right) b - \frac{e}{4 e_* \lambda_K^2} B & = -\zeta_K \frac{\Phi_K}{2\lambda_K^2} \delta(x) \delta(y) .
\end{align}
\label{eq:Bcond}
\end{subequations}
\end{widetext}

These equations are valid for $r = \sqrt{x^2 + y^2} > \xi,\xi_K$. Inside the vortex core or above $T_c$ or $ T_K$, we just replace $\lambda \rightarrow \infty$ and analogously for $\lambda_K$.

\subsubsection{Case $\Delta \neq 0 , V \neq 0$.}

Integration of Eq.~\eqref{eq:Bcond} over space leads to $\Phi_{\rm tot} = \int d^2 x B = B(q = 0), \phi_{\rm tot} = \int d^2x b = b(q = 0)$ and thereby to Eq.~\eqref{eq:Fluxes} of the main text.

We now determine the spatial dependence of $B(r), b(r)$.
The Green's function of Eq.~\eqref{eq:Bcond}

\begin{align}
    G(\v r) &= \frac{1}{2\pi} \left(\begin{array}{cc}
        \frac{K_0(r/2\lambda_K) - (2\lambda_K/\lambda)^2K_0(r/\lambda)}{(1 - (2\lambda_K/\lambda)^2)}& 0 \\
        0 & K_0(r/2\lambda_K)
    \end{array}\right)  \notag\\
    &+ \underbrace{ \frac{K_0(r/\lambda) - K_0(r/2\lambda_K)}{2\pi (1 - (2\lambda_K/\lambda)^2)}}_{=: \mathcal F(r)/2\pi } \left (\begin{array}{cc}
        1 & \frac{e}{e_* } \\
       \frac{e}{ e_*}  & \frac{e^2}{e_*^2}
    \end{array}\right).
    \end{align}
Here, $K_0(r)$ is the zeroth modified Bessel function of the second kind.
We here consider the solution when both order parameters are present. For simplicity, we concentrate on $\text{max}(\xi_K,\xi) < \text{min}(\lambda_K,\lambda)$, which should be valid sufficiently far from $T_c$.

Using this Green's function we find
\begin{align}
\left (\begin{array}{c}
B \\ b
\end{array} \right) & = \frac{1}{2 \pi} \left ( \begin{array}{cc} 
\frac{\zeta \Phi_0}{\lambda^2} K_0 \left (\frac{r}{\lambda} \right)\\
-\frac{2\zeta_K \Phi_K}{(2\lambda_K)^2} K_0 \left (\frac{r}{2\lambda_K} \right)
\end{array}\right)
\notag\\
 &+\frac{1}{2 \pi} \left ( \begin{array}{cc} 
-{2\zeta_K \Phi_0} \left ( \frac{e}{e_*}\right)^2  \frac{K_0 \left (\frac{r}{\lambda} \right)-K_0 \left (\frac{r}{2\lambda_K} \right)\left (\frac{\lambda}{2\lambda_K}\right)^2}{\lambda^2 - (2\lambda_K)^2}\\
\frac{\zeta \Phi_K}{\lambda^2} \mathcal F(r)
\end{array}\right).
\end{align}
Note that in the limit when $\zeta = 0$, the physical $B$ field is of order $e^2/e_*^2 \ll 1$.

\subsubsection{Case $\Delta = 0$ but $V \neq 0$.}

To access the regime above $T_c$ we set $1/\lambda^2 = 0$ in Eq.~\eqref{eq:Bcond}. To obtain the total flux, we use the $q \rightarrow 0$ limit of the Fourier transformed fields

\begin{subequations}
\begin{align}
    [b - e B/e_*]_{q \rightarrow 0}& = - 2 \zeta_K \Phi_K,\\
    [eb/e* + B]_{q \rightarrow 0 } &= 0. 
\end{align}
\end{subequations}
Inverting this pair of equations leads to Eq.~\eqref{eq:Fluxes} of the main text.

We determine the spatially dependent fields in the limit $e^2 \ll e_*^2$ and obtain
\begin{align}
         \left (\begin{array}{c}
        B \\
        b
    \end{array} \right) &= \zeta_K \frac{\Phi_K}{\lambda_K^2} \frac{1}{2} \frac{K_0(r/2\lambda_K)}{2\pi} \left (\begin{array}{c}
       \frac{e}{e_*} \\
       - 1     \end{array}\right). \label{eq:Bb}
\end{align}

We can estimate the magnetic field in the vortex core (the highest field induced by the vortex) by its value at the boundary of the vortex core, i.e.
\begin{equation}
    B_{\rm max} = \Phi_0 \frac{e^2}{e_*^2} \frac{\ln(\kappa_K)}{4\pi \lambda_K^2}. \label{eq:BSM}
\end{equation}
This is the result quoted in the main text below Eq.~\eqref{eq:Fluxes}.

\subsubsection{Case $\Delta \neq 0$ but $V = 0$.}

Finally, from Eq.~\eqref{eq:Bcond} at $1/\lambda_K =0$, we reproduce the textbook results for type-II superconductors for total fluxes $\Phi_{\rm tot} =\zeta \Phi_0$ and fields $B(r)  = \frac{\zeta \Phi_0 K_0(r/\lambda)}{2 \pi \lambda^2}$, while $b(r) =0$. 

\subsection{String energy}
Generalizing textbook arguments to the present case of $V$ and $\Delta$ vortices and using $\mathbf B = B(r) \hat e_z$ and $\mathbf b = b(r) \hat e_z$ we obtain at the saddle point level
\begin{align}
    f_{\rm SC} + f_{\rm K} & = \frac{\lambda^2}{8 \pi} (\partial_r B(r) + e \partial_r b(r)/e_*)^2 +\frac{\lambda_K^2}{2 \pi} (\partial_r b(r) )^2.
    \end{align}
Note that for $T>T_c$ the first term drops from the equation, see Eq.~\eqref{eq:Bb}.   

\subsubsection{Case $\Delta \neq 0, V\neq 0$.}

We thus obtain for the free energy contribution of the string
\begin{widetext}
\begin{align}
    \epsilon_{\rm str}(\zeta, \zeta_K)& \simeq \frac{\Phi_0}{4\pi} \Big \{ \zeta^2 H_\xi \frac{\ln\left(\kappa\right)}{2 \kappa^2} + \zeta_K^2 \frac{e}{e_*} h_{\xi_K} \frac{\ln\left(2\kappa_K\right)}{2 \kappa_K^2} \notag\\
     & + H_\xi \frac{e^2}{2\kappa^2e_*^2} \left [(\zeta - 2 \zeta_K)^2 \frac{e^2}{e_*^2}  + \zeta^2 \frac{(2\lambda_K)^2}{\lambda^2} \right ][I_3(\lambda, 2\lambda_K/\lambda) - I_3(2\lambda_K,2\lambda_K/\lambda)] \notag \\
     & + H_\xi \frac{e^2}{ 2\kappa^2 e_*^2} \left [2 \zeta (\zeta - 2 \zeta_K) I_3(\lambda, 2 \lambda_K/ \lambda) - 4\zeta \zeta_K  I_3(2\lambda_K, 2\lambda_K/\lambda) \right ] \Big \rbrace . \label{eq:Etot}
\end{align}
\end{widetext}

We here use 
\begin{align}
   I_3(\lambda, b) & \simeq - \frac{1}{2 \left(1-b^2\right)}-\frac{\ln (b)}{\left(b^2-1\right)^2},\\
   I_3(2\lambda_K, b) & \simeq \frac{1}{2 \left(1-b^2\right)}+\frac{b^2 \ln (b)}{\left(b^2-1\right)^2}.
\end{align}
Note that the second and third line in Eq.~\eqref{eq:Etot} is negligible for $\lambda \sim \lambda_K$. This is the limit under consideration in Eq.~\eqref{eq:FstringMaintext} of the main text.

\subsubsection{Case $\Delta = 0$ and $V \neq 0$}

Because of the above discussion of the zero mode, only the $f_K$ contributes to the kinetic energy and we obtain

\begin{align}
    \epsilon_{\rm str} 
    & \simeq \frac{\zeta_K^2 \Phi_K}{(4\pi)} h_{\xi_K} \frac{ \ln(2\kappa_K)}{2 \kappa_K^2}.
\end{align}

\subsubsection{Case $\Delta \neq 0$ and $V = 0$.}
In this case
\begin{align}
    \epsilon_{\rm str} & = \frac{\zeta^2 \Phi_0}{(4\pi)} H_\xi \frac{ \ln(\kappa)}{2 \kappa^2}.
\end{align}

\subsection{Ground state free energy}


We consider a magnetic field in $z$ direction and $\mathbf H = H e_z, \mathbf h = h e_z$ and note that $h<0$ will also be considered ($H<0$ is related to our findings by symmetry) and consider only the leading order $H,h$.

\subsubsection{Case $\Delta \neq 0$ and $V \neq 0 $}

We thus find for the free energy functional for vortices $\zeta, \zeta_K$ that the change in free energy per vortex string length is
\begin{equation}
    \frac{\Delta F}{L}  = \epsilon_{\rm str}(\zeta,\zeta_K) - \frac{1}{4\pi} (\zeta \Phi_0 H + (\zeta - 2 \zeta_K) \Phi_K h). \label{eq:FullStringSM}
\end{equation} 

\subsubsection{Cases $\Delta = 0, V \neq 0 $ or $\Delta \neq 0, V = 0$}

In these cases we find
\begin{subequations}
\begin{align}
    \frac{\Delta F}{L}  
    &=    \frac{\Phi_K}{4\pi}\left [\zeta_K^2 h_{\xi_K}\frac{\ln(2\kappa_K)}{2 \kappa^2_K} -  2\zeta_K ( \frac{e}{e_*} H -  h) \right ],\\
    \frac{\Delta F}{L} & = \frac{\zeta \Phi_0}{4\pi} [\zeta H_\xi \frac{\ln(\kappa)}{2\kappa^2} - H ].
\end{align}
\end{subequations}
The upper line corresponds to the condition for the orbitally limited Kondo effect, while the lower line is the condition for $H_{c,1}$ for the SC.

Thus, apart from the some coefficients, the relationship between the definition of the upper critical field and lower critical fields is in parallel to the case of the superconductor and here given by (cf Eq.~\eqref{eq:UpperCritFields})
\begin{subequations}
    \begin{align}
    H - H_\xi \frac{\ln(\kappa)}{2\kappa^2} & = 0,\\
    \vert \frac{e}{e_*} H - h \vert- h_{\xi_K} \frac{\ln(2 \kappa_K)}{(2 \kappa_K)^2} &= 0.
\end{align}
\label{eq:TrivialLowerCritFields}
\end{subequations}

\subsection{Phase diagram}

We here determine the phase diagram in the plane of $H$ and $h$ in Fig.~\ref{fig:MagField}. We use Eqs.~\eqref{eq:UpperCritFields} to determine the upper critical fields, while we use Eqs.~\eqref{eq:TrivialLowerCritFields} to determine the lower critical fields in the case of only one order parameter. We can further numerically determine the phase diagram in the presence of two mean-field order parameters by minimizing the free energy Eq.~\eqref{eq:FullStringSM} over all possibility $\zeta, \zeta_K \in \{-1,0,1\}$. This leads to a phase diagram depicted in Fig.~\ref{fig:MagField} a).

\begin{figure}
    \centering
    \includegraphics{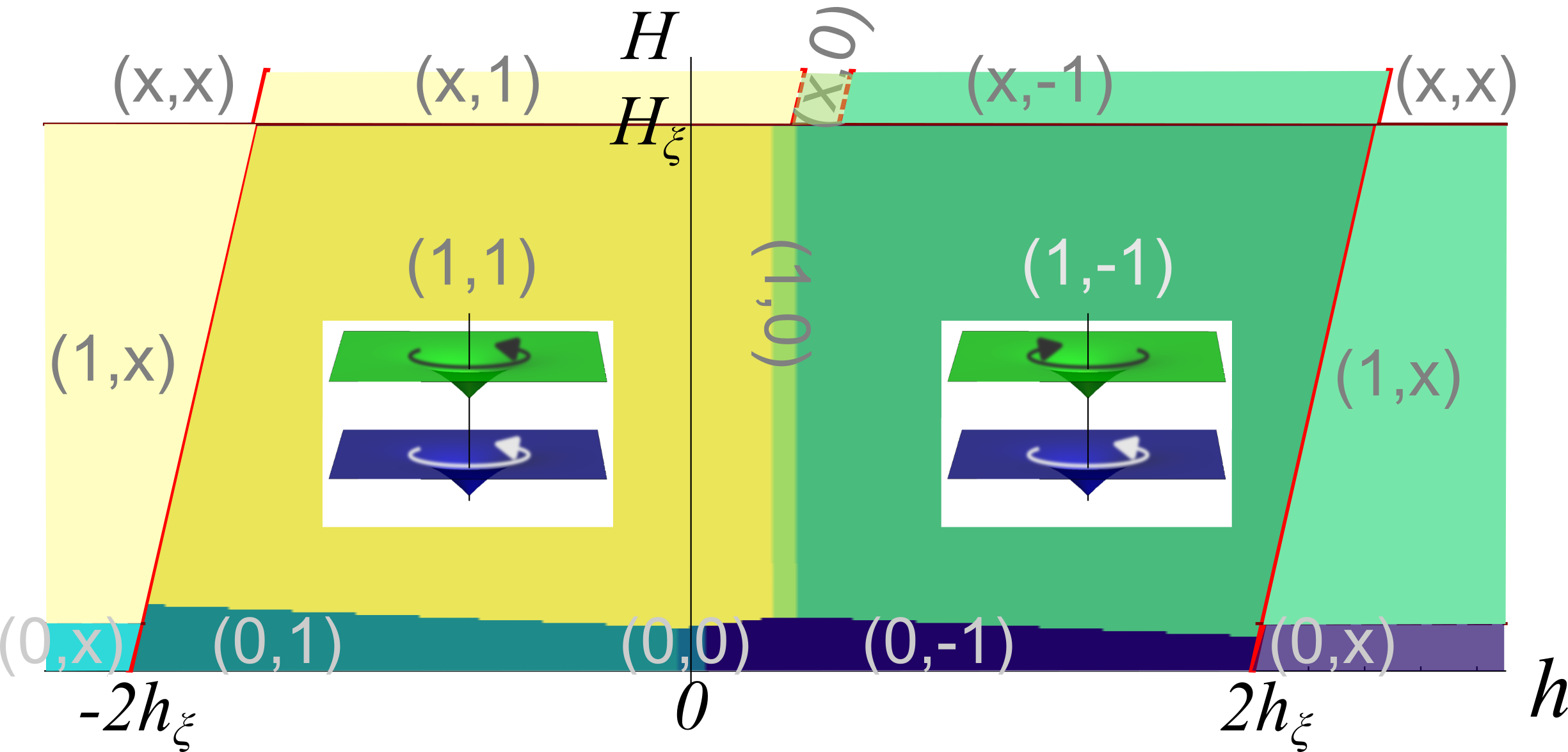}
    \caption{The phase diagram analogous to Fig.~\ref{fig:MagField} of the main text, but for $e^2/e_*^2 = 10^{-2}$, keeping all other parameters equal.}
    \label{fig:MagFieldSM}
\end{figure}

In Fig.~\ref{fig:MagField} a), it appears as if the conditions Eq.~\eqref{eq:TrivialLowerCritFields} trivially held also in the presence of both order parameters. This is not generally true and an artifact of the small $e^2/e^2_*$ limit whilst keeping $\lambda/ \lambda_K$ finite. To illustrate these intricacies, in Fig.~\ref{fig:MagFieldSM} we present the analog of Fig.~\ref{fig:MagField} a) for $e^2/e^2_* = 10^{-2}$ keeping all other parameters the same.

\section{Microscopic derivation of Ginzburg-Landau Parameters}
\label{sec:MicroscopicsSM}

In this section of the supplement we derive the microscopic parameters of the Ginzburg-Landau functional starting from a large-$N$ description of a mixed-valence system, see e.g. Eq. (13.3) of \cite{ColemanBook} supplemented with Heisenberg interaction of spins. This allows to microscopically demonstrate the existence of the type-II heavy Fermi liquid phase, identifying a regime $\lambda_K \gg \xi_K$, at least within the mean-field framework.

We use the identification $V = t_{cd} b /\sqrt{N}$, where $t_{cd}$ is the hopping matrix elements between localized $1-T$ orbitals denoted by $d^\dagger$-creation operators and conduction electrons $c^\dagger$
\begin{align}
S &=  \int d \tau \sum_{\v r} \Big \lbrace c^\dagger_{\v r, \alpha} [\partial_\tau + \lambda_{\v r}]f _{\v r, \alpha} \notag \\
& + f^\dagger_{\v r, \alpha} [\partial_\tau + \lambda_{\v r}]f _{\v r, \alpha} \notag \\
& + \frac{
N}{t_{cd}^2}V^\dagger_{\v r, \alpha} [\partial_\tau + \lambda_{\v r} - E_f]V _{\v r} \notag \\
& + [V_{\v r} c^\dagger_{\v r, \alpha} f_{\v r, \alpha} + H.c.] - (\lambda_{\v r}-E_f)Q \Big \rbrace \notag \\
& +\int d \tau\sum_{ \langle \v r, \v r' \rangle} [ \frac{N \vert t_{\langle\v r, \v r' \rangle }\vert^2}{J_H} -  (f^\dagger_{\v r, \alpha} t_{\langle\v r, \v r' \rangle }f_{\v r', \alpha} + H.c. )]. \label{eq:DecoupledMicroAction}
\end{align} 

Here we assume a separation of scales between the temperatures of the $\langle t_{\langle\v r, \v r' \rangle} \rangle \sim t_f$ condensation (i.e., mean-field QSL behavior) and $V$ condensation and expand in $\vert t_f \vert \gg \vert V\vert$, so that we find on the mean-field level
\begin{equation}
    \frac{t_f}{J_H} = \langle f^\dagger_{\v r} f_{\v r + \hat \epsilon} \rangle + H.c. ,\label{eq:MFtf}
\end{equation}
where $f$-fermion correlators are calculated in the presence of spinon hopping $t_f$. This lead to $t_f \sim J_H$ at lowest temperatures~\cite{AuerbachBook}.

\begin{figure}
    \centering
    \includegraphics[scale = 1]{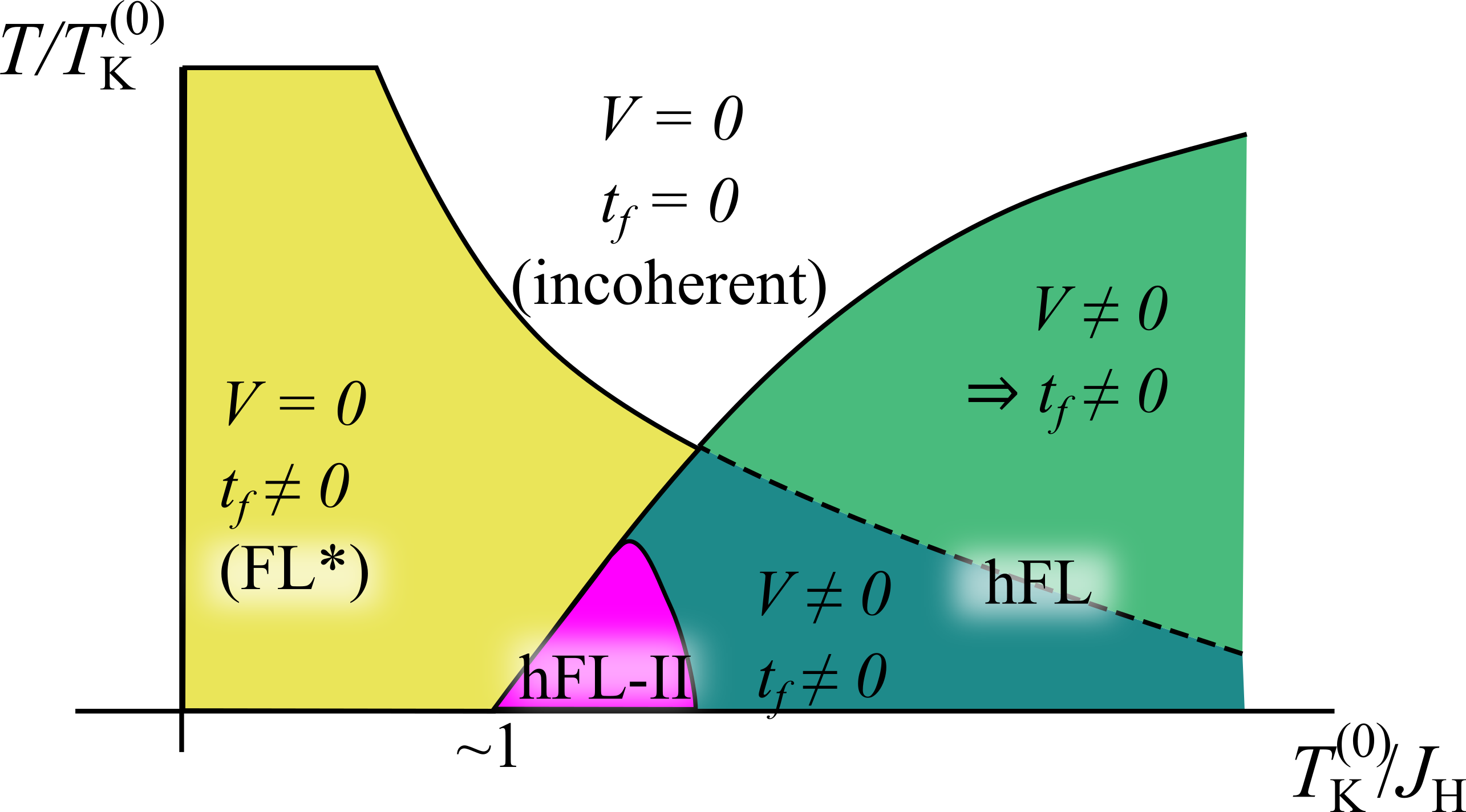}
    \caption{Schematic mean-field phase diagram for Eq.~\eqref{eq:DecoupledMicroAction}. One may expect four phases depending on whether $t_f \neq 0$ or $t_f = 0$ and $V \neq 0$ or $ V = 0$. However, non-zero $V$ implies non-zero $t_f$~\cite{SenthilSachdev2004} (spinon hopping is generated to $\mathcal O(V^2 t_c)$), leaving only three phases. The heavy Fermi liquid (hFL) occurs in two types: The conventional type-I at large $T_K^{(0)} \gg J_H$, and additionally the type-II hFL (hFL-II, in pink). In this supplement, we demonstrate that at least within mean-field theory it does appear near the quantum phase transition at $T_K^{(0)} \sim J_H$. It is expected~\cite{SenthilSachdev2004} that beyond mean-field theory, a quantum critical fan arises around this point, see Fig.~\ref{fig:Summary} c) of the main text.}
    \label{fig:PhaseDiagSuppMat}
\end{figure}

We will consider temperatures near the mean-field U(1) FL$^*$ to hFL transition, where per definition $\vert t_f \vert \gg \vert V\vert$, cf. Fig.~\ref{fig:PhaseDiagSuppMat}. We will also consider Fermi pockets of conduction electrons which are smaller than the spinon Fermi pocket.
For the triangular lattice with nearest neighbor hopping, replacing the Fermi surface at half-filling with a circle is a good approximation. This motivates a parabolic approximation of both conduction electron and spinon bands, i.e. $S = S_{\rm free} + S_{\rm hyb}$

\begin{subequations}
\begin{align}
    S_{\rm free} & = \int d\tau (dp) 
    c_\alpha^\dagger(\tau, \v p) [\partial_\tau - \xi(\v p)] c_\alpha(\tau,\v p) \notag \\
    & +f_\alpha^\dagger(\tau, \v p) [\partial_\tau - \xi_f(\v p)] f_\alpha(\tau,\v p),  \\
    S_{\rm hyb}&= \int d\tau  d^d x N \frac{\vert V(\v r)\vert^2 }{a^2 J_K^{\rm eff}} + [ V c^\dagger_\alpha f_\alpha + H.c.] . 
\end{align}
\label{eq:SfreeAndHyb}
\end{subequations}

Here, we introduced 
\begin{equation}
    J_K^{\rm eff} = \frac{t_{cd}^2}{\lambda - E_f} >0
    \end{equation}
and $\lambda \approx 0.8 t_f$ is given by the dispersion of spinons on the triangular lattice.

For simplicity and to be concrete, we here consider purely 2D motion, the inclusion of a third dimension will only quantitatively affect the results, and we focus on particle-like (hole-like) electrons (spinons) $\xi = \v p ^2/2m - \mu$ ($\xi_f(\v p) = - \v p^2/2m_f + \mu_f$), as well as the limit of $m_f \gg m$, where, effectively, the band masses are $t_c a^2 \sim 1/m$, $t_f a^2 \sim 1/m_f$.
We will further use the notation $\bar m = m m_f/(m_f + m) \equiv m m_f/m_{\rm tot}$ for the reduced mass. We assume a small conduction electron pocket $p_c < p_f \sim 1/a$, where $p_c^2 = 2 m \mu \; (p_f^2 = 2m_f \mu_f)$, but not too small such that $\mu = p_c^2/2m \gg \mu_f = p_f^2/2m_f$.

\subsection{Ginzburg-Landau parameters for homogeneous fields}

We calculate the susceptibility bubble for the hybridization ``order parameter'' [$\int d^2p/(2\pi)^2 = \int_{\v p}$]

\begin{align}
    \chi_K & = T\sum_{\epsilon_n} \int_{\v p} \frac{1}{i \epsilon_n -  \xi(\v p)}\frac{1}{i \epsilon_n - \xi_f(\v p )} \notag \\
    & = \int_{\v p} \frac{\tanh\left ( \frac{\xi_f(\v p)}{2T}\right)-\tanh\left ( \frac{\xi(\v p)}{2T}\right)}{2[\xi(\v p) - \xi_f(\v p)]}.
\end{align}

Note that the pole at $\xi = \xi_f$ is avoided. 
In the limit of small temperatures we obtain
\begin{align}
    \chi_K & = \int_{\v p} \frac{\text{sgn}(\xi_f)\theta(- \xi\xi_f)  + 2 T [\delta (\xi) - \delta (\xi_f )]}{\xi - \xi_f} \notag\\
    & \simeq \frac{ m}{2\pi}\ln \left (\frac{\mu^2}{\mu_f^2}  \frac{1}{(1-p_c^2/p_f^2)^3}\right) - \frac{ T}{\pi\mu_f} \frac{m}{1- p_c^2/p_f^2 }.
\end{align}
Here, we used the that the upper cut-off of the momentum integral is of the same order of magnitude as $p_f$. In the last line, we kept leading terms in the limit of $m_f \gg m$, $\mu \gg \mu_f$. Note that we formally expanded in small temperatures, which is justified for $T< \mu_f (1 - p_c^2/p_f^2)$.
We thus find
\begin{align}
    \alpha_K &= \frac{N}{a_c} [\frac{1}{a^2 J_K^{\rm eff}} - \chi_K] \notag\\
    & \simeq \frac{N m}{\pi a_c} \left [ \ln \left ( \frac{\mu_f (1-p_c^2/p_f^2)^{3/2}}{T_K^{(0)}} \right) + \frac{ T}{\mu_f} \frac{1}{1- p_c^2/p_f^2 } \right ] \notag\\
    & \simeq \frac{N m }{\pi a_c} \frac{\sqrt{1 - p_c^2/p_f^2}(T - T_K)}{T_K^{(0)}} ,
\end{align}
where $a$ ($a_c$) are in-plane (out-of-plane) lattice constants.
We used the following notation 
for the mean-field Kondo temperature 
\begin{align}
T_K &= \frac{T_K^{(0)}}{\sqrt{1 - p_c^2/p_f^2}} - \mu_f (1 - p_c^2/p_f^2) \notag\\
&\ll T_K^{(0)} \equiv \mu e^{-2\pi/J_K^{\rm eff} a^2 m}. 
\end{align}
Note that the mean-field QPT occurs at $\mu_f = T_K^{(0)}/(1-p_c^2/p_f^2)^{3/2}$ and $\mu_f \sim J_H$.

The fourth order term can be written as 
\begin{align}
    \beta_k &= \frac{N}{a_c}T \sum_{\epsilon_n}\int_{\v p}\left ( \frac{1}{i \epsilon_n - \xi(\v p)}\frac{1}{i \epsilon - \xi_f(\v p)} \right)^2  \notag\\
    & = \frac{2 N}{a_c} \int_{\v p}  \frac{\text{sgn}(\xi(\v p)\theta(- \xi(\v p)\xi_f(\v p))}{[\xi(\v p)-\xi_f(\v p)]^3} \notag\\
    & \simeq \frac{N m}{\pi a_c} \frac{1}{ \mu_f^2 \left ( 1 - p_c^2/p_f^2\right)}.
\end{align}

\subsection{Stiffness}

Next we determine the stiffness below $T_K$ and by these means the parameter $m_K$ of the Ginzburg-Landau functional. The stiffness, as obtained by second order gradient expansion of the free energy in $\nabla \phi$ is
\begin{align}
    K_{ij} & = - \frac{2 N V^2}{a_c} T\sum_{\epsilon_n} \int_{\v p} \frac{\partial_{p_i}\xi\partial_{p_j}\xi_f}{[V^2 + (\epsilon_n + i \xi)(\epsilon_n + i \xi_f)]} \notag \\
    & =  \frac{\delta_{ij} N V^2}{a_c} \int_{\v p}\frac{\v p^2}{m m_f} \frac{\theta(V^2 - \xi  \xi_f)}{[4V^2 - (\xi - \xi_f)]^{3/2}} \notag\\
    & \simeq \delta_{ij}\frac{ N V^2}{2\pi a_c \mu_f} \frac{m}{m_f(1 - p_c^2/p_f^2)}.
\end{align}
The last line is not only an expansion in small $m/m_f$, $\mu_f/\mu$.
This implies for the parameter of the kinetic term in the Ginzburg-Landau functional that
\begin{equation}
    \frac{1}{2m_K} = \frac{N}{2\pi a_c \mu_f} \frac{m}{m_f} \frac{1}{(1-p_c^2/p_f^2)} 
    = \frac{N m}{\pi a_c}\frac{a^2}{(1 - p_c^2/p_f^2)} .
\end{equation}
This leads to the coherence length
\begin{equation}
    \xi_K = a \sqrt{\frac{\mu_f}{2(T_K- T)}} \frac{1}{(1-p_c^2/p_f^2)^{1/4}}, \label{eq:CohLengthSM}
\end{equation}
or in, in units of $V= \sqrt{\vert \alpha_K \vert/\beta_K}$
\begin{align}
    \xi_K &= 1/\sqrt{2m_K \beta_K V^2 }  \notag\\
    & = a \frac{\mu_f}{V}  \label{eq:CohLengthSM2}.
\end{align}
Note that the Kondo coherence length is thus given by the smaller of the two velocities divided by the order parameter.

\subsection{Ring exchange terms}

Finally, we estimate the effect of ring-exchange terms on the mean-field level following Ref.~\cite{Motrunich2006}. These additional magnetic interaction terms stemming from ring exchange have the form 
\begin{widetext}
\begin{align}
H_{\rm ring} &= - \sum_{\v r_{1,2,3} \in \triangle} \frac{J_3}{N^2} f_{ABC }f^\dagger_{\v r_1,\alpha_1}f_{\v r_1,\beta_1}f^\dagger_{\v r_2,\alpha_2}f_{\v r_2,\beta_2}f^\dagger_{\v r_3,\alpha_3}f_{\v r_3,\beta_3}  \tau^A_{\alpha_1 \beta _1} \tau^B_{\alpha_2 \beta _2} \tau^C_{\alpha_3 \beta _3} \notag\\
&- \sum_{\v r_{1,2,3,4} \in \lozenge} \frac{J_4}{N^3} \chi_{ABCD }f^\dagger_{\v r_1,\alpha_1}f_{\v r_1,\beta_1}f^\dagger_{\v r_2,\alpha_2}f_{\v r_2,\beta_2}f^\dagger_{\v r_3,\alpha_3}f_{\v r_3,\beta_3} f^\dagger_{\v r_4,\alpha_4}f_{\v r_4,\beta_4}  \tau^A_{\alpha_1 \beta _1} \tau^B_{\alpha_2 \beta _2} \tau^C_{\alpha_3 \beta _3}\tau^D_{\alpha_4 \beta _4}.
\end{align}
Here, $f_{ABC} = \tr[\tau^A \tau^B \tau^C]$ (fully antisymmetrized with respect to all indices) and $\chi_{ABC} = \tr[\tau^A \tau^B \tau^C \tau^D]$ (fully symmetrized for all indices), and $\tau^A$ are the generators of SU(N) (traceless, hermitian, N$\times$N matrices). For $N>2$, there is also a term 
\begin{equation}
    \delta H_{\rm ring} = - \sum_{\v r_{1,2,3} \in \triangle} \frac{J_3'}{N^2} d_{ABC }f^\dagger_{\v r_1,\alpha_1}f_{\v r_1,\beta_1}f^\dagger_{\v r_2,\alpha_2}f_{\v r_2,\beta_2}f^\dagger_{\v r_3,\alpha_3}f_{\v r_3,\beta_3}  \tau^A_{\alpha_1 \beta _1} \tau^B_{\alpha_2 \beta _2} \tau^C_{\alpha_3 \beta _3},
\end{equation}
\end{widetext}
where $d_{ABC} = \tr[\tau^A \tau^B \tau^C]$ fully symmetrized over indices. We can use for SU(N) generators $f_{ABC} f_{ABC} = N(N^2 - 1) \sim N^3$, $d_{ABC} d_{ABC} = (N^2-4)(N^2 - 1)/N \sim N^3$  and $\chi_{ABCD}\chi_{ABCD}  = [2(N^4 - 1) + (N^2-4)(N^2+2)(N^2 - 1)]/24N^2 \sim N^4$, as well as $\langle f^\dagger_{\v x} f_{\v x + \hat e} \rangle = t_f/2J_H \simeq 1$, so that at the mean-field level and at large $N$\cite{Motrunich2006,KoenigKomijani2021}
\begin{equation}
    F_{\rm ring}/N \sim  J_3 \sin(\phi) - \left [J_3'  + J_4/24 \right ] \cos(\phi).
\end{equation}
Here, $\phi =\oint d \vec l \cdot d \tilde{\v a}(\v x)$ is the flux enclosed in a unit cell in a gauge in which we write the spinon hopping term $f^\dagger ( - i \nabla - \tilde{\v a}) f$. Identification $e_* \v a  = \tilde{\v a}$ and $\phi = e_* a^2 {\v b}_z$ leads to the free energy Eq.~\eqref{eq:GL} as in the main text with 
\begin{equation}
    \frac{1}{e_*^2}  \sim N \frac{a^2}{a_c} J_4
\end{equation}    
    and $ \vert \v h \vert  \sim N J_3 e_*/a_c$. For simplicity we have dropped $J_3'$ terms vanishing in the SU(2) limit (note that $J_3$ and $J_4$ terms survive the $N = 2$ limit). We also remark that in the limit of large $J_4 \sim J_H$ one may estimate $e_*^{-2}\sim v_f$ (the speed of spinons), corresponding to a dimensionless coupling of order unity, as expected for U(1) QSL with spinon Fermi surface. 

\subsection{Order of magnitude of important scales}

This concludes the estimate of the ratio of London penetration and coherence lengths as quoted in the main text.
We can estimate their ratio as
\begin{align}
    \kappa_K & = \sqrt{\frac{m_K^2 \beta_K    }{e_*^2}} \notag \\
    & \simeq \sqrt{\frac{ J_4 t_c}{\mu_f^2}(1- p_c^2/p_f^2)}.
\end{align}
Note that within our calculation one may replace $\mu_f \sim J_H$ (and near the Mott transition $J_4 \sim J_H$) and $p_f \sim 1/a$. Note that a $T_K^{(0)}/J_H$ larger than the QPT value, additional factors of $t/J_H$, Eq.~\eqref{eq:MFtf}, enter and reduce $\kappa_K$ such that it eventually drops below unity.

Using the coherence length, Eq.~\eqref{eq:CohLengthSM}, we obtain for the Kondo-London penetration depth
\begin{equation}
    \lambda_K = a\sqrt{ \frac{J_4 t_c}{2\mu_f (T_K-T)}} (1-p_c^2/p_f^2)^{1/4},
\end{equation}
which for $J_4 \sim \mu_F \sim J_H$ leads to the rough estimate $\lambda_K \sim a \sqrt{ 
 t_c/(T-T_K)} \sim .1 \mu \text{m}/(1-T/T_K)^{1/2}$.
We here use $a \sim 1$ nm in the CDW phase, $T_K \sim 1$K (if we identify the maximal temperature for the magnetic memory with the mean-field Kondo temperature) and $t_c \sim 10000 $K. 

For the estimate of the magnetic field in the vortex core, which relies on Eq.~\eqref{eq:BSM}, we additionally employ (using $a_c \sim a$)
\begin{equation}
    \frac{e^2}{e_*^2} = \alpha_{\rm QED} \frac{J_4 a}{c},
\end{equation}
so that 
\begin{align}
    B_{\rm max}&= \Phi_0 \alpha_{\rm QED}\frac{1}{c} \frac{\ln(\kappa_K)}{4\pi a} \frac{2 \mu_f (T_K-T)}{ t_c \sqrt{1 - p_c^2/p_f^2}}  \notag \\
    & = \underbrace{\frac{\Phi_0}{a^2}}_{\sim 10^3 \text{T}} \underbrace{\frac{a \alpha_{\rm QED} T_K}{2 \pi c}}_{\sim 10^{-9}} \frac{m}{m_f} \frac{\ln(\kappa_K) (1-T/T_K)}{\sqrt{1- p_c^2/p_f^2}}.
\end{align}
In the main text, we drop the last term, assuming $p_c \lesssim p_f, T \lesssim T_K$ and $\ln(\kappa_K) = \mathcal O(1)$. This concludes the estimate presented below Eq.~\eqref{eq:Fluxes} of the main text.

\section{Tunneling in the vortex core}
\label{sec:CdM}

In this section of the supplement, we discuss the tunneling density of states above the superconducting transition temperature. 
We concentrate on the center of the vortex and, to be concrete, employ the same model of parabolic bands, Eq.~\eqref{eq:SfreeAndHyb}, as in the microscopic derivation of the Ginzburg-Landau free energy, except that we now fix $V(\v r) = V(r) e^{i \phi}$ to contain a vortex.

\subsection{Subgap states}

At the mean-field level, the first quantized fermionic Hamiltonian is thus
\begin{equation}
h = \left (\begin{array}{cc}
\frac{\v p^2}{2m} - \mu & V(\v r) \\ 
V^*(\v r) & -[\frac{\v p^2}{2m_f} - \mu_f]
\end{array} \right).
\end{equation}

Here, the first and second row/column correspond to $c$ and $f$ electrons, respectively.
The interesting regime is $V \ll t_f \sim 1/m_f a^2 \ll t_c \sim 1/m a^2$ and energies close to the gap opening, yet our calculations are generic with respect to the ratio $t_c/t_f$ (the $t_c = t_f$, $\mu = \mu_f$ is equivalent to a superconductor). 
We concentrate on the energy of the band opening, measure all energies with respect to this energy and use that at this energy $k_c = \sqrt{2m \mu}$ and $k_f = \sqrt{2 m_f \mu_f}$ are equal (denoted by $k_{\rm gap}$). 
The Schr\"odinger equation in polar coordinates is
\begin{widetext}
\begin{equation}
\left (\begin{array}{cc}
-\frac{1}{2m} [\frac{1}{r} \partial_r (r \partial_r \bullet )+ \frac{\partial_\phi^2}{r^2}] - \mu & V(r) e^{i \phi} \\ 
V(r) e^{-i \phi} & \frac{1}{2m_f} [\frac{1}{r} \partial_r (r \partial_r \bullet )+ \frac{\partial_\phi^2}{r^2}] + \mu_f
\end{array} \right) \left (\begin{array}{c}
\psi_c(r, \phi) \\ 
\psi_f(r, \phi)
\end{array} \right) = E \left (\begin{array}{c}
\psi_c(r, \phi) \\ 
\psi_f(r, \phi)
\end{array} \right).
\end{equation}

In the following, we closely follow the classical work~\cite{CaroliMatricon1964} by Caroli, deGennes, and Matricon (but generalize it to $m \neq m_f, \mu \neq \mu_f$) and use Pauli matrices $\sigma_{x,y,z}$ for $c/f$ space. We parametrize
$\left (\psi_c(r, \phi),
\psi_f(r, \phi) \right )^T
= e^{i l \phi - i \phi \sigma_z/2} \hat \phi_l(r) $
leading to 
\begin{equation}
\left (\begin{array}{cc}
\frac{1}{2m} [-\frac{1}{r} \partial_r (r \partial_r \bullet )+ \frac{(l - 1/2)^2}{r^2}] - \mu & V(r) \\ 
V(r)& -\frac{1}{2m_f} [-\frac{1}{r} \partial_r (r \partial_r \bullet )+ \frac{(l + 1/2)^2}{r^2}] + \mu_f
\end{array} \right) \hat \phi_l = E \hat \phi_l.
\end{equation}
\end{widetext}
 
Note that $l$ is quantized to half-integers to make $\psi_{c,f}(r, \phi)$ single-valued.  
Following Caroli, de Gennes and Matricon, we define a radius $r_c \in [l/k_{\rm gap}, \xi_K]$, where $\xi_K$ is the radius of the vortex (determined by the Kondo coherence length, Eq.~\eqref{eq:CohLengthSM2}).

\subsubsection{Solution deep inside the core: $r < r_c$}

We first find the solution in the regime $r \ll r_c$ deep inside the vortex core. Clearly $V(r) \simeq 0$ and positive $l$ solutions are
\begin{equation}
\hat \phi_l(r) = \left (\begin{array}{c}
A_c J_{l - 1/2 }([k_{c}  + E/v_c] r)\\
A_f J_{l + 1/2 }([k_{f}  - E/v_f] r)
\end{array}\right),
\end{equation}
while negative $l$ solutions are by symmetry
\begin{equation}
\hat \phi_l(r) = \left (\begin{array}{c}
A_c J_{\vert l \vert + 1/2 }([k_{c}  + E/v_c] r)\\
A_f J_{\vert l \vert - 1/2 }([k_{f}  - E/v_f] r)
\end{array}\right).
\end{equation}

We remind the reader that only $J_0(1) = 1$, all other Bessel functions vanish at the origin. Thus, only the $l = 1/2$ state has support of $c-$electrons at the origin.

\subsubsection{Solution for $r> r_c$}
Next we search for the solution outside for $r >r_c$, note that this still includes part of the outer vortex core. The Ansatz
\begin{equation}
\hat \phi_l(r) = \hat g_l(r) H_m(k_{\rm gap} r) + c.c.,
\end{equation}
 
where $\hat g_l(r)$ is slow on the scale $1/k_{\rm gap}$ and $m = \sqrt{l^2 + 1/4}$. We obtain

\begin{equation}
\left (\begin{array}{cc}
-i v_c  \partial_r   & V(r) \\ 
V(r)& i v_f \partial_r 
\end{array} \right) \hat g_l  =  \left (\begin{array}{cc}
 E + \frac{l}{2mr^2} & 0\\ 
0 & E +  \frac{l}{2m_f r^2}
\end{array} \right)   \hat g_l. \label{eq:ZMEq}
\end{equation}
 Here $v_c = k_{\rm gap}/m$ and $v_f = k_{\rm gap}/m_f$ and their difference and geometric average are called $\Delta v$, $\bar v = \sqrt{v_c v_f}$.

We generalize the Ansatz of Caroli, deGennes and Matricon and write 
\begin{equation}
\hat g_l(r)  = A e^{-K(r)} \left (\begin{array}{cc} e^{i \psi_c/2\bar v} & 0\\ 0 & e^{- i \psi_f/2\bar v}\end{array}\right ) {\hat g}^{(0)}.
\end{equation}
Here, $K(r)  =  \int_0^{r} dr' V(r')/\bar v$, $A$ is a normalization constant and 
\begin{equation}
{\hat g}^{(0)}  = \frac{1}{\sqrt{v_c + v_f}}\left ( \begin{array}{c}
\sqrt{v_f} \\ - i \sqrt{v_c}
\end{array}\right),
\end{equation}
is the zero mode of the operator on the left of the equation sign in Eq.~\eqref{eq:ZMEq}.

Using the result for ${\hat g}^{(0)}$ we obtain two equations
\begin{subequations}
\begin{align}
\sqrt{\frac{v_c}{v_f}} [i V(r) + \psi_c' - i e^{- i (\psi_c + \psi_f)/2\bar v} V(r)] & =  E + \frac{l}{2m r^2} , \\
\sqrt{\frac{v_f}{v_c}} [-i V(r) + \psi_f' + i e^{ i (\psi_c + \psi_f)/2\bar v} V(r)] & = E + \frac{l}{2m_f r^2}.
\end{align}
\end{subequations}
Next we use the notation $\psi_\pm = \frac{\psi_c \pm \psi_f}{2}$ to and expand in $\psi_+$ to obtain
\begin{subequations}
\begin{align}
[\partial_r - V(r)/\bar v] \psi_+ & = \bar v \left[\frac{E(v_f + v_c)}{2\bar v^2} + \frac{l}{2k_{\rm gap}r ^2} \right ], \\
\partial_r  \psi_- & =  E\frac{v_f - v_c}{\bar v},
\end{align}
\end{subequations}

which has solutions 
\begin{subequations}
\begin{align}
\psi_+ &=-\bar v\int_r^\infty dr' e^{K(r)-K(r')}  \left[\frac{E(v_f + v_c)}{2\bar v^2} + \frac{l}{2k_{\rm gap}{r '}^2}\right ],\\
\psi_- & = E(v_f-v_c)r/\bar v + \text{const.} .
\end{align}
\end{subequations}

We briefly check that  $\psi_+/\bar v$ is small. For large $r \gg \xi \sim v_c/V_\infty > \bar v/V_\infty$ we can use 
\begin{align}
\psi_+(r)/\bar v &\simeq
-\left [\frac{E(v_f + v_c)}{2 V_\infty \bar v} + \frac{l}{2k_{\rm gap} r}\right ],
\end{align}
which is small so long as $r k_{\rm gap} \gg 1$.
For $r \ll \xi_K$ we can simplify
\begin{subequations}
\begin{align}
K(r) &=  \int_0^r dr' V(r')/ \bar v\sim r^2/\xi_K^2\\
\psi_+ & \simeq \frac{E (v_f + v_c) r}{2 \bar v} \notag \\
& -\bar v\int_0^\infty dr' e^{-K(r')}  \left[\frac{E(v_f + v_c)}{2\bar v^2} - \frac{l V(r')}{2k_{\rm gap}\bar v r'}\right ].
\end{align}
\end{subequations}
The first term is much smaller than $\bar v$ by means of $E \ll \mu_f$, while the second term turns out to be fixed to zero by the quantization condition (see below).

In total we find for $r \geq r_c$
\begin{equation}
\hat \psi_l = A H_m(k_{\rm gap} r) \frac{e^{-K(r)}}{\sqrt{v_c + v_f}}\left ( \begin{array}{c}
e^{i \frac{\psi_+ + \psi_-}{2 \bar v}} \sqrt{v_f}\\
-i e^{i \frac{\psi_+ - \psi_-}{2 \bar v}} \sqrt{v_c}\\
\end{array} \right) + c.c. . \label{eq:PsiLarger}
\end{equation}

\subsubsection{Matching solutions}

Finally, we match the solution obtained at small and large $r$. The
Expansion of Bessel functions for positive $l$ for the solution deep in the core leads to 
\begin{align}
\hat \psi_l(r_c) 
&\simeq  \sqrt{\frac{2}{\pi k_{\rm gap} r}}\left (\begin{array}{c}
A_c \cos \left ( [k_{\rm gap} + E/v_c]r_c - \frac{l \pi}{2} \right)\\
A_f \sin \left ( [k_{\rm gap} - E/v_f] r_c - \frac{l \pi}{2}\right)
\end{array}\right)
\end{align}

At the same time, the asymptotic form of Hankel functions implies for Eq.~\eqref{eq:PsiLarger}

\begin{widetext}
\begin{align}
\hat \psi_l(r_c) 
&\simeq A \sqrt{\frac{2}{\pi k_{\rm gap} r_c}} e^{i \left (k_{\rm gap} r_c - \frac{m \pi}{2} - \frac{\pi}{4}\right)} \frac{e^{-i\int_0^\infty dr' e^{-K(r')}  \left[\frac{E(v_f + v_c)}{2\bar v^2} - \frac{l V(r')}{2k_{\rm gap}\bar v r'}\right ]}}{\sqrt{v_c + v_f}} \left ( \begin{array}{c}
e^{i \frac{Er_c}{v_c}} \sqrt{v_f} \\
-i e^{-i \frac{Er_c}{v_f}} \sqrt{v_c} 
\end{array} \right) + c.c.
\end{align}
\end{widetext}

By comparison of exponents we obtain the quantization condition becomes

\begin{equation}
\int_0^\infty dr' e^{-K(r')}  \left[\frac{E(v_f + v_c)}{2\bar v} - \frac{l V(r')}{2k_{\rm gap} r'}\right ] =0,
\end{equation}
or simply
\begin{equation}
{E_l = \frac{2\bar v^2}{v_f + v_c}  \frac{l}{2k_{\rm gap} \bar v} \frac{\int_0^\infty dr' e^{-K(r')}  \frac{V(r')}{r'}}{\int_0^\infty dr' e^{-K(r')}  }}.
\end{equation}
Using the minigap energy $E_{\rm mg} = \bar v V'(0)/2k_{\rm gap} (v_f + v_c) \simeq V_{\infty}^2/\sqrt{v_c v_f k_{\rm gap}^2} $ we thereby find the result quoted in the main text. We highlight that the (indirect) spectral gap is $V_{\rm gap} = 2\sqrt{v_f/v_c} V_\infty$ (see below), hence 
\begin{equation}
    E_{\rm mg}  =\frac{ V_{\rm gap}^2}{\mu_f} \sqrt{\frac{v_c}{v_f}}.
\end{equation}

Note that the comparison of prefactors $\vert A_c/A_f\vert  \propto \sqrt{v_f/v_c} = \sqrt{m/m_f}$ implies that the c-electron component of the subgap states is strongly suppressed.

\subsection{Tunneling spectroscopy}

Microscopically, the tunneling Hamiltonian is 
\begin{equation}
    H_{\rm tun} = \sum_{\v x, \v x'} p^\dagger_\alpha(\v x) [ t_{pc}(\v x, \v x')c_{\alpha}(\v x') +t_{pd}(\v x, \v x')d_{\alpha}(\v x') ] + H.c.
\end{equation}

For the spatially resolved tunneling density of states near the center of the vortex this implies in the slave-boson framework

\begin{align}
    \frac{d\text{I}}{d\text{V}}(\v x) &\propto \int d^2 x' \sum_l \delta(e\text{V} - E_l) \notag \\
    & \times\left \vert (t_{pc}(\v x, \v x'), V(\v x') \frac{t_{pd}(\v x, \v x')}{t_{cd}}) \left ( \begin{array}{c}
         \psi_c(\v x')  \\
         \psi_f(\v x')
    \end{array}\right )\right \vert^2.
\end{align}

We consider the limit of a relatively sharp tip, $t_{pc, pd}(\v x, \v x')  \simeq 0$ for $\vert \v x - \v x' \vert > r_{\rm tip}$ and $r_{\rm tip} \ll \xi_K$. We thus find at the center of the vortex
\begin{align}
     \frac{d\text{I}}{d\text{V}} &\propto \sum_{l>0} \delta(e\text{V} - l E_{\rm mg}) \int_0^{k_{\rm gap} r_{\rm tip}} d\rho \rho\notag \\
    & \times\left \vert t_{pc} \sqrt{v_f} J_{l - 1/2}(\rho) + \frac{V_\infty}{t_{cd}} t_{pd} \frac{\rho}{k_{\rm gap}\xi_K} \sqrt{v_c} J_{l + 1/2}( \rho)\right \vert^2 \notag \\
    & + \sum_{l<0} \delta(e\text{V} - l E_{\rm mg}) \int_0^{k_{\rm gap} r_{\rm tip}} d\rho \rho\notag \\
    & \times\left \vert t_{pc} \sqrt{v_f} J_{\vert l \vert + 1/2}(\rho) + \frac{V_\infty}{t_{cd}} t_{pd} \frac{\rho}{k_{\rm gap}\xi_K} \sqrt{v_c} J_{\vert l \vert - 1/2}( \rho)\right \vert^2.
\end{align}

We define 
\begin{align}
    R & \equiv \frac{V_\infty t_{pd}}{t_{cd}t_{pc}} \sqrt{\frac{v_c}{v_f}} \frac{1}{k_{\rm gap} \xi_K} \notag\\
    & \simeq \frac{V_\infty^2 t_{pd}}{t_{cd}t_{pc}t_f} \sqrt{\frac{\mu}{t_f}}  \sim \frac{T_K^2 }{t_{cd}^{1/2}t_f^{3/2}} \frac{t_{pd}}{t_{pc}},
\end{align}
where in the last line we assumed $t_{cd} \sim \mu$. At least on the 1H surface where $t_{pd} < t_{pc}$ it is valid to assume that $R \ll 1$, and it appears plausible that this assumption also holds on the 1T surface if at least some electrons tunnel through the 1T layer into the 1H layers.

Assuming that $k_{\rm gap} r_{\rm tip} \sim 1$, the relative strength of $c$-tunneling, mixed tunneling and $f$-tunneling is thus
\begin{align}
    1 : R :R^2,
\end{align}
and by consequence 
\begin{align}
     \frac{d\text{I}}{d\text{V}} &\propto \sum_{l} \delta(eV - l E_{\rm mg}) \int_0^{k_{\rm gap} r_{\rm tip}} d\rho \rho J_{\vert l \vert - \text{sign}(l)1/2}(\rho). \label{eq:TDOSCDGM}
\end{align}
For the plot in the main text, Fig.~\ref{fig:MagField} b), c), we use $k_{\rm gap} r_{\rm tip} = 3$, use a Lorentzian broadening of the delta function

\begin{equation}
    \delta(e\text{V} - l E_{\rm mg}) \rightarrow \frac{\Gamma}{(e\text{V} - l E_{\rm mg})^2 + \Gamma^2},
\end{equation}
where $\Gamma = 0.2 E_{\rm mg}$. 

\subsection{Tunneling outside the vortex core}

For clarity of presentation, we overlay the density of states in the vortex core with the same observable outside the vortex core using the homogeneous Hamiltonian 
\begin{equation}
    h = \left (\begin{array}{cc}
        v_c p & V  \\
        V & -v_f p
    \end{array}\right).
\end{equation}
Note the spectral gap is $V \sqrt{1 - \left ( \frac{v_c + v_f}{v_c - v_f}\right)^2} \simeq 2\sqrt{v_f/v_c} V$. The tunneling density of states is
\begin{align}
    \frac{d \text{I}}{d\text{V}} &\propto \frac{v_c}{v_c + v_f} \Big [ \left (1 + \frac{t_{pd}^2 V^2}{t_{pc}^2t_{cd}^2} \right)I_1 \left (eV/V, \frac{v_c - v_f}{v_c + v_f} \right ) \notag \\
   &+\frac{2t_{pd} V}{t_{pc} t_{cd}} I_2 \left (eV/V, \frac{v_c - v_f}{v_c + v_f} \right )\Big ] , \label{eq:TDOSBULK}
\end{align}

where
\begin{subequations}
    \begin{align}
        I_1(z, \alpha) &= \sum_{\pm} \int_{-\infty}^\infty dx \delta(z - E_\pm(x)) \\
        I_2(z, \alpha) &= \sum_{\pm} \pm \int_{-\infty}^\infty dx \delta(z - E_\pm(x)) \frac{1}{\sqrt{x^2 + 1}}
    \end{align}
\end{subequations}
and $E_\pm(x) = \alpha x \pm \sqrt{x^2 + 1}$. In Fig.~\ref{fig:MagField} c), d) we plot Eq.~\eqref{eq:TDOSBULK} for $t_{pc} = 0.1 V,
t_{pd} = 0.1 V, 
t_{cd} = 2V. 
v_f = .05 V$.
Given that we only calculated the proportionality constant of the tunneling density of states for both subgap states, Eq.~\eqref{eq:TDOSCDGM} and far away from vortex cores \eqref{eq:TDOSBULK}, the relative strength of bulk (green) and subgap contributions (blue) are not to scale. 

\end{document}